%% file: main.tex
\documentclass[10pt,twocolumn,letterpaper]{article}

\usepackage[pagenumbers]{iccv} 

\usepackage{multirow}
\usepackage{tablefootnote}
\usepackage{amsmath,bm}
\usepackage{graphicx}
\usepackage{subcaption}

\definecolor{refblue}{rgb}{0.21,0.49,0.74}
\usepackage[pagebackref,breaklinks,colorlinks,allcolors=refblue]{hyperref}
\usepackage{makecell}

\title{DreamCube: 3D Panorama Generation via Multi-plane Synchronization}

\author{
Yukun Huang$^{1}$,
Yanning Zhou$^{2}$,
Jianan Wang$^{3}$,
Kaiyi Huang$^{1}$,
Xihui Liu$^{1\dagger}$
\vspace{4pt}\\
$^{1}$The University of Hong Kong ~ $^{2}$Tencent ~ $^{3}$Astribot\\
\vspace{-10pt}\\
{\url{https://yukun-huang.github.io/DreamCube/}}
}

\begin{document}


\twocolumn[{
\maketitle
\vspace{-0.3cm}
\centerline{
\includegraphics[width=\linewidth,trim={4pt 4pt 4pt 4pt}]{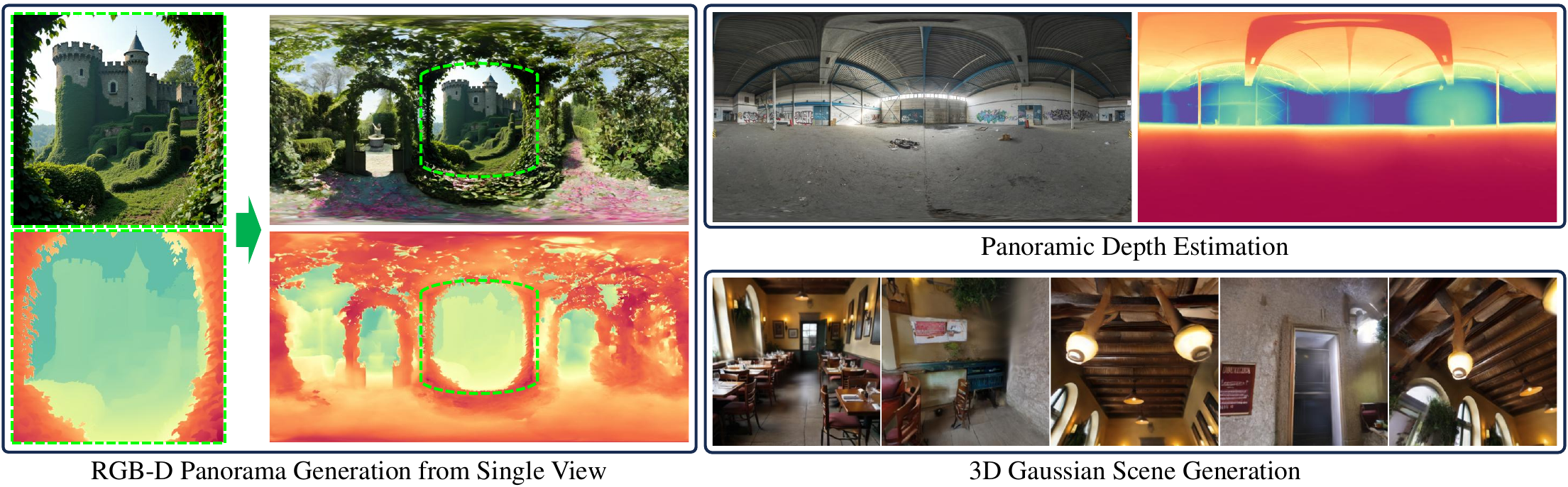}}
\vspace{-0.1cm}
\captionof{figure}{In this work, we introduce \textbf{Multi-plane Synchronization} to generalize 2D diffusion models to multi-plane omnidirectional representations (\ie, cubemaps), and \textbf{DreamCube} for RGB-D cubemap generation. The proposed approaches can be applied to different tasks including RGB-D panorama generation, panorama depth estimation, and 3D scene generation.}
\label{fig:teaser}
\vspace{3em}
}]

\renewcommand{\thefootnote}{}
\footnotetext{$\dagger$ Corresponding author.}
\renewcommand{\thefootnote}{\arabic{footnote}}

\input{sections/0_abstract}
\input{sections/1_introduction}
\input{sections/2_related_work}
\input{sections/3_method}
\input{sections/4_experiment}
\input{sections/5_conclusion}
{
    \small
    \bibliographystyle{ieeenat_fullname}
    \bibliography{reference}
}

\end{document}

%% file: sections/0_abstract.tex
\begin{abstract}
3D panorama synthesis is a promising yet challenging task that demands high-quality and diverse visual appearance and geometry of the generated omnidirectional content. Existing methods leverage rich image priors from pre-trained 2D foundation models to circumvent the scarcity of 3D panoramic data, but the incompatibility between 3D panoramas and 2D single views limits their effectiveness. In this work, we demonstrate that by applying multi-plane synchronization to the operators from 2D foundation models, their capabilities can be seamlessly extended to the omnidirectional domain. Based on this design, we further introduce DreamCube, a multi-plane RGB-D diffusion model for 3D panorama generation, which maximizes the reuse of 2D foundation model priors to achieve diverse appearances and accurate geometry while maintaining multi-view consistency. Extensive experiments demonstrate the effectiveness of our approach in panoramic image generation, panoramic depth estimation, and 3D scene generation.
\end{abstract}

%% file: sections/1_introduction.tex
\section{Introduction}
\label{sec:intro}

Humans live in a fully immersive 3D environment. Simulating this immersive experience is crucial for applications such as virtual reality, gaming, and robotics ~\cite{somanath2021hdr,yang2024dreamspace,zhu2024point}.
As a fundamental technology for building 3D world, omnidirectional content synthesis aims to generate visual content that covers a full $360^\circ \times 180^\circ$ field of view, encompassing both appearance and geometry.
Despite this necessity, modern generative models~\cite{latentdiffusion,flux2024,esser2024scaling} require large amounts of training data, yet the scale of currently available omnidirectional assets remains relatively small compared to conventional perspective images. 

Leveraging the rich image priors from pre-trained 2D diffusion models~\cite{latentdiffusion}, previous works~\cite{panodiff,diffpano,panfusion,diffusion360,cubediff,mvdiffusion,panogen,stitchdiffusion,omnidreamer} explore repurposing diffusion-based image generators to create $360^\circ$ panoramas, which circumvents the problem of insufficient panoramic data.
While most of these works~\cite{panodiff,diffpano,panfusion,diffusion360,stitchdiffusion,omnidreamer} adopt equirectangular projection (ERP) for panoramas for simplicity, this approach introduces significant challenges.
The ERP causes severe spatial distortions near the poles, resulting in a pixel distribution fundamentally different from that of perspective images on which the pre-trained models were trained.
These distortions not only affect visual quality but also limit the transferability of pre-trained knowledge, resulting in suboptimal generation quality particularly at the poles. 

\begin{figure}[tbp]
\centering
\includegraphics[width=\linewidth]{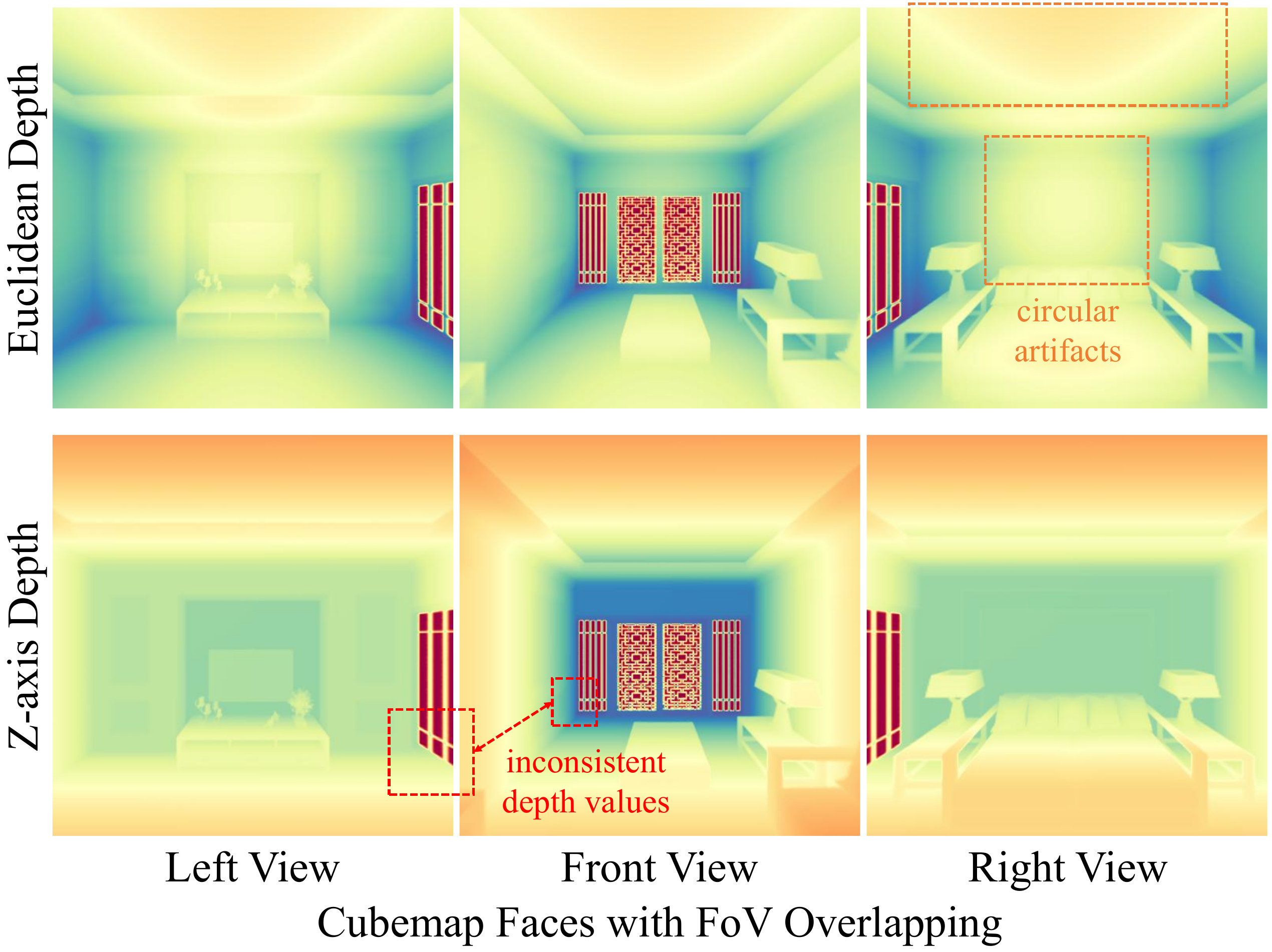}
\caption{\textbf{Motivation.} Previous works~\cite{ldm3d,panodiffusion} on RGB-D panorama generation is based on equirectangular representations and only supports Euclidean depth instead of the more popular Z-depth. However, the distribution of Euclidean depth is quite different from that of RGB images (\eg, circles on flat surfaces as highlighted by \textcolor{orange}{\textbf{orange}} dashed boxes), which hinders the use of pre-trained 2D diffusion priors. Multi-plane methods~\cite{cubediff,mvdiffusion} support Z-depth, but their adopted FoV overlapping techniques lead to depth ambiguity, as highlighted by \textcolor{red}{\textbf{red}} dashed boxes. Different from existing works, our approach supports multi-plane Z-depth generation without using FoV overlapping techniques.}
\label{fig:depth}
\end{figure}

Another solution for panoramic synthesis is the multi-plane approach, which utilizes 2D diffusion to generate synchronized multi-perspective images. In general, multi-plane images are closer to the in-domain distribution of the pre-trained 2D diffusion models and can natively produce higher resolution panoramas. However, the separate generation of multiple planes leads to severe seam inconsistencies. To the best of our knowledge, all existing works~\cite{mvdiffusion,cubediff} on multi-plane panorama generation adopt field-of-view (FoV) overlapping to improve seam inconsistencies. Although effective to some extent, the overlapping planes actually cause computational redundancy and reduce the effective image resolution. Moreover, we empirically found that FoV-overlapping multi-planes are exhibit significant incompatibilities in non-image domains, such as the latent space of LDM~\cite{latentdiffusion} and the Z-depth domain, as shown in Figure~\ref{fig:depth}. 
These incompatibilities manifest as conflicting, non-unique z-depth values at overlapping regions, leading to the artifacts in the end.
These limitations lead to a fundamental question: \textit{Can seam inconsistencies be resolved without FoV overlapping?}

\begin{table}[!t]
    \begin{center}
    \caption{Comparisons of different panorama generation methods. Omni. refers to Omnidirectional (360°$\times$ 180°).}
    \footnotesize
    \renewcommand{\arraystretch}{1.2}
    \setlength{\tabcolsep}{3pt}  

    \begin{tabular}{ccccc}
        \hline
        \makecell{Methods} & \makecell{Condition} & \makecell{Output} & \makecell{Resolution$^\dagger$} & \makecell{Omni.}\\
        \hline
        Text2Light~\cite{text2light} & text & HDR & $512\times1024$ & \checkmark \\ 
        LDM3D-Pano~\cite{ldm3dvr} & text & RGB-D & $512\times1024$ & \checkmark  \\
        OmniDreamer~\cite{omnidreamer} & image & RGB & $256\times512$ & \checkmark  \\
        Diffusion360~\cite{diffusion360} & text/image & RGB & $512\times1024$  & \checkmark  \\
        MVDiffusion~\cite{mvdiffusion} & text/text-image & RGB & $256\times256\times8$ $^{\ddagger}$& $\times$ \\
        PanFusion~\cite{panfusion} & text/text-layout & RGB & $512\times1024$& \checkmark  \\
        PanoDiffusion~\cite{panodiffusion} & image & RGB-D &$256\times512$  & \checkmark  \\
        CubeDiff~\cite{cubediff} & text-image & RGB &  $491\times491\times6$ $^{\ddagger}$& \checkmark  \\
        \hline
         DreamCube (Ours) &  text-image & RGB-D &$512\times512\times6$ & \checkmark  \\
        \hline
    \end{tabular}
        \label{tab:panorama_methods}

    \end{center}
    \vspace{-0.3cm}
\footnotesize{$^\dagger$Base resolution without super-resolution or post-processing.}\\
\footnotesize{$^{\ddagger}$Calculated effective resolution accounting for overlapping fields of view.}
\end{table}

To address this question, we present the first thorough analysis of 2D diffusion models to identify the causes of seam inconsistencies in multi-plane generation.
Our analysis reveals that these inconsistencies are rooted in the translational inequivalence of certain neural operators in the omnidirectional image domain. 
Surprisingly, we find that by adapting these operators to be omnidirectionally translation-equivalent, existing 2D diffusion models can generate seam-consistent panoramic multi-planes without requiring fine-tuning or FoV overlapping. In this work, we refer to this adaptation as ``multi-plane synchronization''.

Building on multi-plane synchronization, we introduce DreamCube, a novel framework for generating RGB-Depth (RGB-D) cube maps from a single view through joint panoramic appearance and geometry modeling. 
Inspired by previous work on diffusion-based depth estimation~\cite{geowizard,lotus,marigold,orchid}, we repurpose the pre-trained 2D diffusion model for multi-plane image and depth generation. 
Unlike previous approaches~\cite{ldm3dvr,panodiffusion} to equirectangular RGB-D panorama generation, we adopt cube maps as panorama representation.
This choice is significant because pixels in each plane of a cube map are uniformly distributed (due to perspective projection rather than equirectangular projection) and therefore align with the in-domain distribution of the pre-trained 2D diffusion models. 
Table~\ref{tab:panorama_methods} compares our approach with existing panorama generation methods across key dimensions.
Together with multi-plane synchronization, our method maximally exploits pre-trained 2D diffusion models for joint modeling of panoramic appearance and geometry. 
The resulting RGB-D cubemaps can be easily lifted to 3D scene, effectively enabling single-view to omnidirectional 3D scene generation.

Our main contributions are as follows:
\begin{itemize}

    \item We thoroughly analyze the operator incompatibilities of existing 2D diffusion models for panoramic multi-plane generation, and propose a multi-plane synchronization strategy that enables seam-consistent cubemap generation without fine-tuning or FoV overlapping.

    \item Based on multi-plane synchronization, we further introduce DreamCube, a masked RGB-D cubemap generator from single view for panoramic appearance and geometry generation.

    \item Extensive experiments demonstrate the effectiveness of our approach in RGB-D panorama generation, panoramic depth estimation, and 3D scene generation.

\end{itemize}

%% file: sections/2_related_work.tex
\section{Related Work}

\noindent \textbf{2D panorama generation.} 
Panorama image generation aims to create 360-degree panoramas from text prompts or partial images. Early works~\cite{cubegan,text2light} utilized GAN~\cite{gan} for panorama generation, while recent diffusion models have enabled more sophisticated panoramic synthesis. 
Current approaches either iteratively outpaint 360° panoramas from narrow field-of-view images~\cite{panogen,multidiffusion,panofree} or directly generate complete panoramas~\cite{panfusion,panodiffusion,mvdiffusion,stitchdiffusion,diffpano,panodiff}.
Many methods employ equirectangular projection (ERP) to represent 360°×180° views, but face challenges with geometric distortions, particularly in polar regions. PanFusion~\cite{panfusion} addresses this through a dual-branch diffusion model combining panorama and perspective domains with specialized attention mechanisms, while DiffPano~\cite{diffpano} introduces a spherical epipolar-aware attention module for consistency. Various denoising and decoding strategies~\cite{panodiffusion,stitchdiffusion,panodiff} have been proposed to reduce stitching artifacts.
To avoid polar distortion and leverage pretrained perspective image diffusion models, alternative approaches~\cite{mvdiffusion,cubediff} generate multiple perspective views. CubeDiff~\cite{cubediff} introduces a cubemap representation that projects the 360° view onto six faces of a cube, effectively eliminating severe distortions while maintaining consistency.
Yet these multi-plane approaches use FoV overlapping to reduce seam artifacts, which introduces computational redundancy and limits effective resolution.

\noindent \textbf{Generative depth estimation.}
Various methods explore diffusion models for depth estimation~\cite{ddp,diffusiondepth,depthgen,ddvm,vpd}. Marigold~\cite{marigold} converts pretrained Latent Diffusion Models into image-conditioned depth estimators. However, directly applying these depth estimation methods to panoramic images presents challenges: the domain gap between ERP and perspective image representations degrades performance. Alternatively, estimating depth for multiple perspective views independently requires an additional alignment step to ensure consistency.
DAC~\cite{depthanycamera} attempts to jointly learn depth in both ERP and perspective spaces within a unified framework, but its effectiveness remains limited.

\noindent \textbf{Joint appearance and geometry modeling.} 
Another line of works jointly models appearance and geometry~\cite{eigen2015predicting,li2015depth,xu2018pad}. GeoWizard~\cite{geowizard} extends Stable Diffusion with a geometry switcher and scene distribution decoupler to jointly predict depth and normals. Orchid~\cite{orchid} trains a VAE for a new joint latent space incorporating depth and normals.
This joint modeling strategy has also demonstrated effectiveness in specific domains, such as human-centric scenarios~\cite{hyperhuman,sapiens,ji2025joint}.
Following this trend, we jointly model appearance and depth information for panorama generation to enhance scene structure understanding and obtain high-quality depth maps for subsequent 3D scene creation.
Unlike previous RGB-D panorama generation methods~\cite{ldm3dvr,panodiffusion}, our approach operates directly on perspective images, better leveraging prior knowledge from pretrained models while jointly modeling appearance and depth for high-quality panoramic scene creation.

%% file: sections/3_method.tex
\section{Multi-plane Synchronization}

\subsection{Preliminary: Panorama Representations}
{Panorama representations} refer to methods of storing 360-degree images or videos, allowing viewers to look in all directions from a single viewpoint.
Equirectangular and cube map formats are the two primary formats due to their compatibility with existing image processing frameworks.

\textbf{Equirectangular} representations project the panoramic content onto a rectangular grid, where the horizontal axis represents the longitude and the vertical axis represents the latitude. This format is one of the most commonly used for storing and transmitting 360-degree images and videos due to its simplicity and compatibility with standard image and video formats. Despite its ease of use, the equirectangular representation suffers from significant distortion, especially near the poles, where the pixels are stretched.

\textbf{Cube map} representations divide the panoramic scene into six square faces of a cube, each representing a 90-degree field of view. The primary advantage of the cube map is its uniform distribution of pixels across the entire field of view, which minimizes distortion compared to the spherical representation. However, it can introduce discontinuities at the edges where the cube faces meet, which may require additional processing to ensure seamless transitions.
In our work, we adopt the cube map representation for its minimal distortion and compatibility with pre-trained 2D diffusion models.

\subsection{Multi-plane Synchronization}\label{sec:sync}

\begin{figure*}[tbp]
\centering
\begin{subfigure}{0.40\linewidth}
\includegraphics[height=8.5cm]{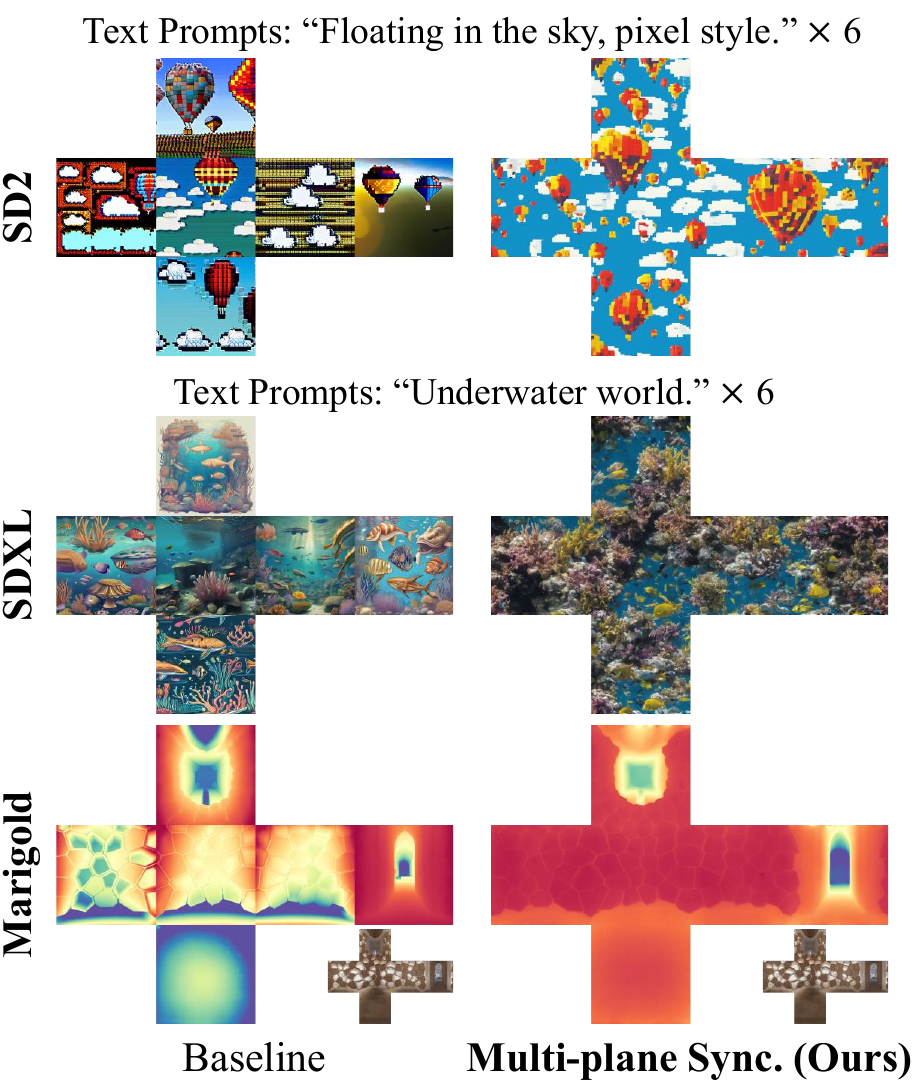}
\caption{Generated multi-plane images.}
\label{fig:sync_demo_a}
\end{subfigure}
\quad\quad
\begin{subfigure}{0.55\linewidth}
\includegraphics[height=8.5cm]{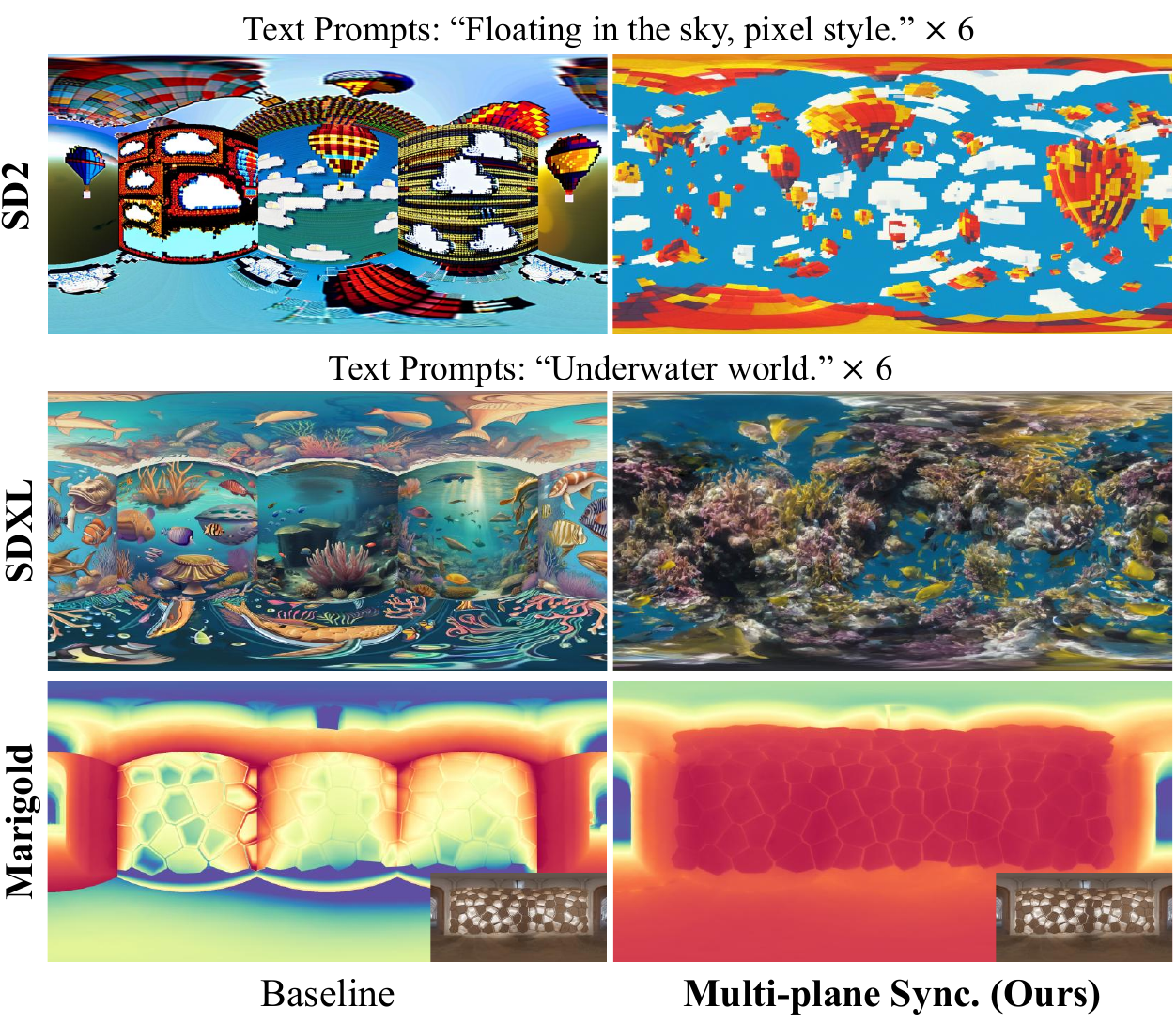}
\caption{Converted equirectangular images.}
\label{fig:sync_demo_b}
\end{subfigure}
\caption{Results of \textbf{Multi-plane Synchronization} on pre-trained 2D diffusion models: SD2~\cite{latentdiffusion}, SDXL~\cite{sdxl}, and Marigold~\cite{marigold}. Our method enables 2D diffusion to generate multi-plane synchronized omnidirectional image representations without fine-tuning.}
\label{fig:sync_demo}
\end{figure*}

Directly applying 2D diffusion models pre-trained on single-view images to multi-plane panoramic representations like cube maps faces a fundamental limitation: They generate each face independently with no inherent correlation, which leads to discontinuities at the seams of adjacent cube faces.
To address this issue, we propose Multi-plane Synchronization, which achieves seamless panoramic generation without the need for fine-tuning or explicitly constructing overlapping regions~\cite{mvdiffusion, cubediff}.

To present our multi-plane synchronization, we use U-Net-based iffusion models~\cite{latentdiffusion} and six-plane cube maps throughout this work. 
The principles established, however, are architecture-agnostic and can be adapted to other diffusion frameworks, such as DiT~\cite{dit} and alternative panorama representation with minimal modifications.

\textbf{Analysis of spatial operators.} 
Neural operators should ideally maintain translation-equivalence in omnidirectional representations, but standard operators (\eg, self-attentions) in 2D diffusion models break this property.
For instance, in standard 2D convolutions, boundary pixels of each cube face are padded with zeros instead of information from adjacent faces, resulting in discontinuities at cube map seams. 
Therefore, all operators of 2D diffusion involving the spatial domain must be adapted to ensure translation invariance in the omnidirectional domain.

\textbf{Synchronizing spatial operators to cube maps.} 
For U-Net-based diffusion models, we adapt three key operators from single-view ($H \times W$) to multi-plane domains ($M \times H \times W$, where $M=6$ for cube maps): (1) synced attentions - reshaping tokens from $(BM)\times(HW)\times C$ to $B\times(MHW)\times C$, enabling attention to operate across all faces simultaneously; (2) synced 2D convolutions - replacing zero-padding with geometrically projected pixels from adjacent faces; and (3) synced group normalizations - calculating statistics globally across all planes rather than per-view independently.

\textbf{Remarks.} 
While individual adaptations of these operators exist in previous works-flattened attentions in~\cite{mvdream,cubediff}, projective resampling in ~\cite{cubegan}, and tiled group normalizations in~\cite{cubediff,scalecrafter}, none provides a comprehensive solution.
Our proposed multi-plane synchronization offers a more complete and systematic adaptation of 2D diffusion models to support the omnidirectional image domain.
As shown in Figure~\ref{fig:sync_demo}, our method can be applied to various pre-trained 2D diffusion models, such as Stable Diffusion~\cite{latentdiffusion} and Marigold~\cite{marigold}, to achieve omnidirectional generation. These results demonstrate the effectiveness and potential of our method in omnidirectional vision tasks such as panorama generation and panoramic depth estimation.

\section{Method}

\begin{figure*}[tbp]
    \centering
    \includegraphics[width=\linewidth]{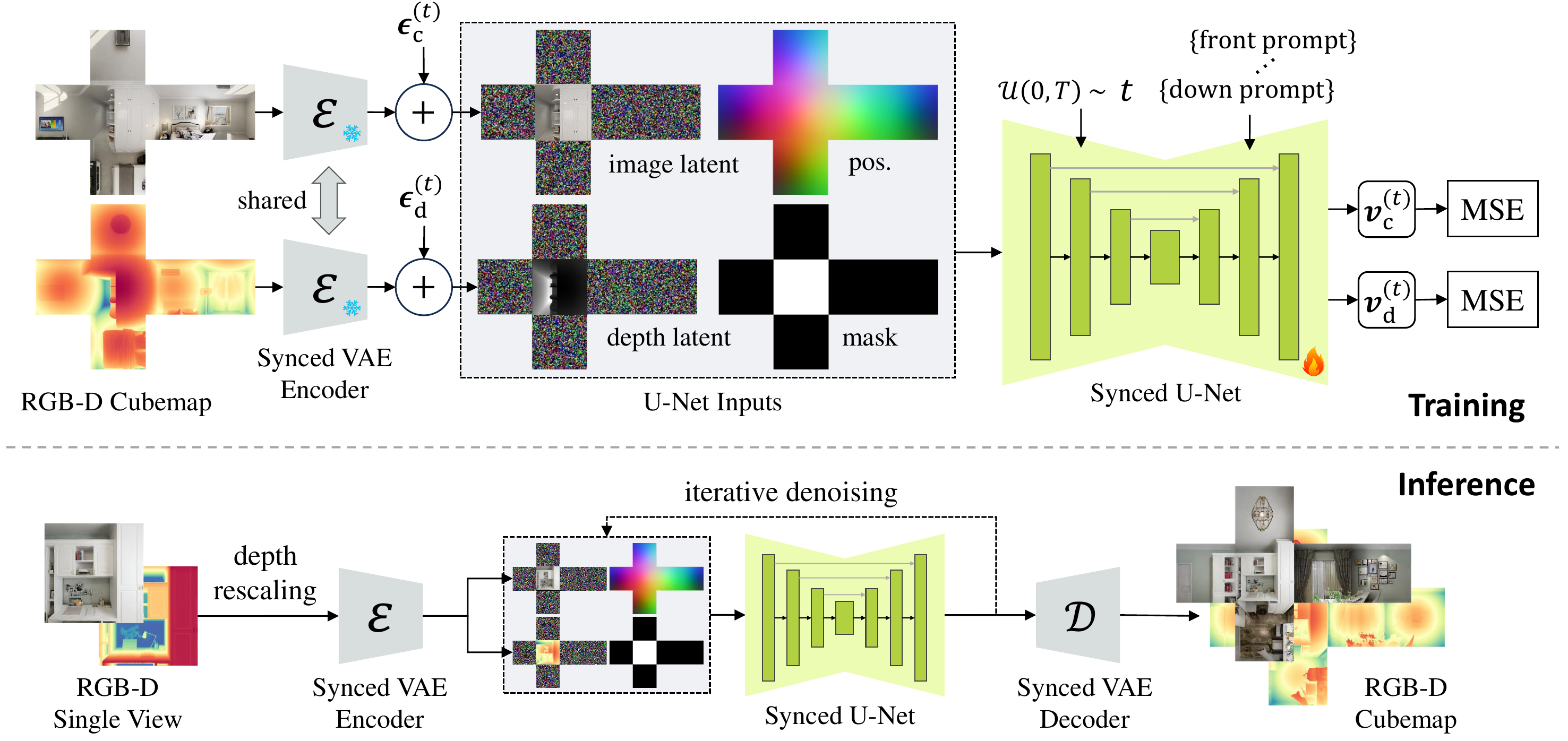}
    \caption{Training and inference framework of \textbf{DreamCube} for RGB-D cube map generation. At training time, RGB-D cube faces are encoded by synced VAE and injected masked Gaussian noises, obtaining image and depth latents. These latents are concatenated with positional encoding and mask as diffusion U-Net's input. The entire U-Net is fine-tuned with $v$-objective~\cite{vprediction} to learn to jointly denoise RGB and depth latents. At inference time, DreamCube receives single-view RGB-D images and multi-view texts as input and generates completed RGB-D cube map representations via iterative diffusion denoising and synced VAE decoding.}
    \label{fig:dreamcube}
\end{figure*}

In this section, we introduce \textbf{DreamCube}, a masked RGB-D cube map diffusion model for joint panoramic appearance and geometry generation.

\subsection{Generative Formulation}
We conceptualize the generation of RGB-D cube map as the production of six colored images $\bm{x}_{c} \in \mathbb{R}^{M \times H \times W \times 3}$ and their corresponding depth maps $\bm{x}_{d} \in \mathbb{R}^{M \times H \times W}$, where $M = 6$ denotes the number of cubic views. Each of these views is associated with a specific face of the cube map, namely \textit{front}, \textit{right}, \textit{back}, \textit{left}, \textit{up}, and \textit{down}. Given single-view RGB-D images and multi-view texts as conditions, our model aims to generate RGB-D images of other cubic views through a synchronous diffusion denoising process.

At training time, we first encode the colored cube map $\bm{x}_{c}$ and the corresponding depth cube map $\bm{x}_{d}$ into latents $\bm{z}_c \in \mathbb{R}^{M \times H' \times W' \times 4}$ and $\bm{z}_d \in \mathbb{R}^{M \times H' \times W' \times 4}$ using a pre-trained VAE~\cite{latentdiffusion}. In particular, the depth inputs $\bm{x}_{d}$ are broadcasted to 3-channels to match the configuration of the VAE. Then, the masked Gaussian noises $\bm{\epsilon}_c^{(t)}$ and $\bm{\epsilon}_d^{(t)}$ are sampled and injected into $\bm{z}_c$ and $\bm{z}_d$, obtaining $\bm{z}_c^{(t)}$ and $\bm{z}_d^{(t)}$, respectively. Note that the conditional image latent (\textit{w.l.o.g.}, we choose the front view as condition) are kept noise-free throughout the diffusion process. Finally, the image latents $\bm{z}_c^{(t)}$ and depth latents $\bm{z}_d^{(t)}$ are concatenated with the positional encoding and mask on the channel axis and fed into the diffusion U-Net. We use $v$-prediction~\cite{vprediction} as the learning objective, and the training loss is given as follows:
\begin{align*}
    \mathcal{L} &= \mathbb{E}_{\bm{x}_c,\bm{\epsilon}_c\sim\mathcal{N}(0, I),t\sim\mathcal{U}(T)} \| \bm{v}^{(t)}_{c} - \hat{\bm{v}}^{(t)}_{c} \|_2^2 \\
     &+ \mathbb{E}_{\bm{x}_d,\bm{\epsilon}_d\sim\mathcal{N}(0, I),t\sim\mathcal{U}(T)} \| \bm{v}^{(t)}_{d} - \hat{\bm{v}}^{(t)}_{d} \|_2^2,
\end{align*}
where $\bm{v}^{(t)}_{c}$ and $\bm{v}^{(t)}_{d}$ are predicted by the diffusion U-Net.

At inference time, the conditional image latents $\bm{c}_c \in \mathbb{R}^{1 \times H' \times W' \times 4}$ and depth latents $\bm{c}_d \in \mathbb{R}^{1 \times H' \times W' \times 4}$ are concatenated with Gaussian noises to get the initial noisy latents $\bm{z}_c^{(T)}$ and $\bm{z}_d^{(T)}$. These noisy latents will be iteratively denoised as the time step decreases from $T$ to $1$ to get noise-free cube map latents $\bm{z}_c^{(0)}$ and $\bm{z}_d^{(0)}$. The final results can be obtained by decoding these latents with VAE.

\subsection{DreamCube: RGB-Depth Cube Diffusion}

As illustrated in Figure~\ref{fig:dreamcube}, DreamCube adopts several designs to enable RGB-D cube map generation from a single view, which involves depth data processing, omnidirectional position encoding, and multi-plane synchronization.

\textbf{Depth data processing.} Panoramic depth data usually uses the Euclidean distance metric, the depth distribution of which is significantly different from the RGB image distribution (\eg, the circle on the flat wall as shown in Figure~\ref{fig:depth}), which hinders the joint modeling of RGB images and depth maps. Therefore, unlike previous works~\cite{panodiffusion,ldm3dvr}, DreamCube models the Z-distance, which is more consistent with the diffusion image priors, rather than the Euclidean distance.

Additionally, DreamCube estimates the depth maps of other cubic views based on the conditional view. Even though the conditional depth values lie in the diffusion's in-domain range \([-1.0, +1.0]\), the generated depths of other views may exceed this range, leading to performance degradation. Inspired by recent work on depth inpainting~\cite{depthlab}, we adopt a depth rescaling strategy. Specifically, we rescale the conditional depth to the range of \([-s, +s]\) before input, where the real number \( s \) is less than 1. This strategy creates an additional margin for the depth generation of other views, thus avoiding out-of-domain depth values.

\begin{figure}[tbp]
\centering
\includegraphics[width=\linewidth]{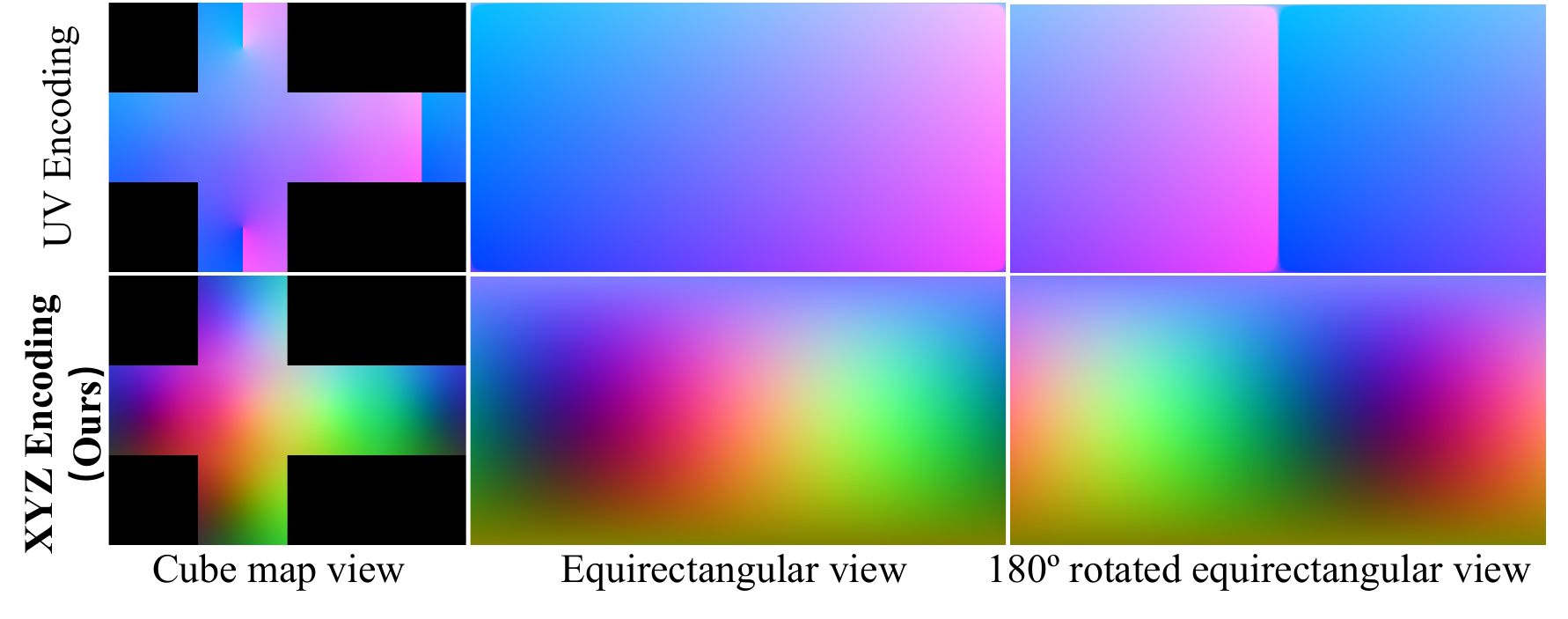}
\caption{Comparison between UV encoding and XYZ encoding for omnidirectional image representations.}
\label{fig:pos_encoding}
\end{figure}

\textbf{Omnidirectional positional encoding.}
To ensure geometric consistency and coherent object relationships across generated cube faces, we improve the spatial awareness of LDM~\cite{latentdiffusion} by integrating positional encodings derived from the 3D geometry of the cube. Specifically, for each point on a cube face, we project its coordinates onto the unit sphere with $(x,y,z)$ values normalized to  $\left[-1, 1\right]$. These normalized coordinates are then appended as three additional channels to LDM's latent representations. This encoding strategy encodes spatial information for each latent patch relative to its cube face while ensuring smooth omnidirectional transitions - addressing limitations of the UV-based encoding proposed by CubeDiff~\cite{cubediff} (see comparisons in Fig.~\ref{fig:pos_encoding}).

\textbf{Multi-plane synchronized diffusion.} As described in Sec.~\ref{sec:sync}, the proposed multi-plane synchronization can improve 2D diffusion models to handle omnidirectional image representations. 
Specifically, all self-attentions, 2D convolutions, and group norms in the diffusion U-Net and VAE are changed to multi-plane synchronized operators. This strategy enables DreamCube to model multi-plane omnidirectional representations, significantly alleviating the tricky issues of discontinuous seams and inconsistent color tones in cube map generation.

\subsection{Building 3D Scene from RGB-D Cubemap}
The generated RGB-D panoramas contain the direction and distance of each pixel, so a colored 3D point cloud can be obtained by projecting all pixels into 3D space. We can further convert the point cloud into different 3D representations, such as 3D meshes and 3D Gaussians~\cite{3dgs}. Note that these conversions can be achieved either by differentiable optimization or by handcrafted algorithms. We choose handcrafted algorithms for fast 3D scene reconstruction in seconds from RGB-D panoramas. Specifically, for 3D mesh reconstruction, we use the obtained point cloud as vertices, and the vertex colors are assigned by the RGB values of the corresponding pixels. The connections between vertices can be extracted based on the adjacency relationship of image pixels. For 3D Gaussians, the position and color of each Gaussian point can be directly assigned from the colored point cloud, while other properties are inferred using a method similar to WonderWorld~\cite{wonderworld}.

It is worth mentioning that different panoramic projections affect the quality of the reconstructed 3D scene. For equirectangular-based RGB-D panoramas, due to the significant geometric distortion at the poles, the reconstructed 3D point cloud will be unevenly distributed and particularly dense at the top and bottom poles. In contrast, the distribution of 3D points from RGB-D cubemap is more uniform and regular, as shown in Figure~\ref{fig:pano2scene_comp}.

%% file: sections/4_experiment.tex
\section{Experiment}

\subsection{Implementation Details}
We implement both Multi-plane Synchronization and DreamCube using PyTorch. For DreamCube, we utilize Stable Diffusion v2~\cite{latentdiffusion} as pre-trained backbone. At training time, we adopt the DDPM noise scheduler~\cite{ddpm} with 1000 timesteps. We use a batch size of $4$ for training, where the resolution of RGB images and depth maps is $512 \times 512$. Random rotation and flipping are used to expand the amount and diversity of panorama training data. The depth rescaling parameter $s$ is randomly sampled from a uniform distribution in $[0.2, 1.0]$ during training. We froze the VAE and fine-tuned the diffusion U-Net for $10$ epochs. We use the AdamW optimizer with a learning rate of $2 \cdot 10 ^{-5}$. The entire training process took approximately two days on four Nvidia L40S GPUs. At inference time, we adopt the DDIM noise scheduler~\cite{ddim} with 50 sampling steps. The depth rescaling parameter $s$ is fixed to 0.6 for inference.

\subsection{Datasets}
We conduct experiments in two distinct data settings to comprehensively evaluate the performance of our method across various scenarios:

\textbf{Indoor setting.} To ensure fair comparison with prior work, we first evaluate our approach on the Structured3D dataset~\cite{structured3d}, which provides equirectangular indoor RGB-D panorama with a $512\times1024$ resolution. Following the same experimental protocol as PanoDiffusion~\cite{panodiffusion}, we utilize their exact data split consisting of 16,930 training, 2,116 validation, and 2,117 test instances. Besides, the SUN360 dataset~\cite{xiao2012recognizing} is also used for out-of-domain evaluation.

\textbf{General setting.} To further evaluate our model's generalization capabilities across diverse environments, we construct a more comprehensive dataset by combining multiple publicly available sources, including Structured3D~\cite{structured3d}, Pano360~\cite{pano360}, Polyhaven~\cite{polyhaven}, Humus~\cite{humus}, HDRI-Skies~\cite{hdriskies} and iHDRI~\cite{ihdri}. This combined dataset encompasses a broad spectrum of both indoor and outdoor environments, resulting in more than 30,000 panoramic instances. This general setting allows us to evaluate the robustness of our approach across a wider range of scenarios.

\textbf{Data processing pipeline.} All panorama data needs to be processed into a unified format, including RGB cube maps, depth cube maps, and image captions for each cube face. For datasets not originally in cubemap format, we apply standard perspective projection to produce cube maps. Next, we adopt BLIP-2~\cite{blip2} to obtain image captions of all cube faces. While the Structured3D dataset includes depth data, the other datasets only contain RGB data. To annotate the depth of these panoramas, we build a high-resolution panorama depth estimation pipeline by connecting the existing panorama depth estimation work Depth Anywhere~\cite{depthanywhere} and the image-guided depth up-sampling work PromptDA~\cite{promptda}, which supports panoramic depth estimation at 4K resolution. We use this pipeline to perform depth estimation on equirectangular-based panoramas and then project the obtained depth panoramas into cube maps.

\subsection{RGB-D Panorama Generation}\label{sec:eval_rgbd}

Evaluating RGB-D panorama generation from a single view presents unique challenges due to the absence of standardized benchmarks.
We train and test our method on the standard split of the Structured3D~\cite{structured3d} dataset following another RGB-D panorama generation work, PanoDiffusion~\cite{panodiffusion}. 
To comprehensively evaluate the capabilities of our method, we evaluate the RGB panorama quality and depth panorama accuracy separately tailored to each modality's characteristics.

\textbf{Evaluation protocol for RGB panorama generation.} We evaluate our method on both in-domain Structured3D~\cite{structured3d} and out-of-domain SUN360~\cite{xiao2012recognizing} datasets. SUN360 consists of around 1000 panorama images including both indoor and outdoor scenes. We use Fréchet Inception Distance (FID)~\cite{heusel2017gans} and Inception Score (IS)~\cite{salimans2016improved} to evaluate the visual quality of generated panorama images.

\textbf{Evaluation protocol for depth panorama generation.} 
Since no ground-truth depths exist for generated panoramas, we propose a reference-based evaluation protocol.
We first project our generated RGB-D panoramas into multiple perspective views at randomly sampled viewpoints.
For each projected RGB image, we obtain a reference depth map using Depth Anything v2~\cite{depthanythingv2}, a state-of-the-art monocular depth estimator.
This provides pseudo ground-truth depth for each perspective view.
We then compare projected depth maps against these reference depths using standard metrics: $\delta$-1.25, AbsREL, RMSE and MAE, following the implementation in~\cite{cheng2018depth}.

\textbf{Quantitative results for RGB panorama generation.} We compare our approach with state-of-the-art panorama generation methods including OmniDreamer~\cite{omnidreamer}, LDM3D-Pano~\cite{ldm3d}, Diffusion360~\cite{diffusion360}, MVDiffusion~\cite{mvdiffusion}, PanoDiffusion~\cite{panodiffusion}, and PanFusion~\cite{panfusion}. The comparison results are reported in Table~\ref{tab:comp_rgb}. Our method performs competitively on both in-domain and out-of-domain datasets with the best overall ranking. Note that for SUN360, we only use it for out-of-domain evaluation while Diffusion360 and OmniDreamer use it for training. Nonetheless, our method significantly outperforms both methods on inception score.

\begin{table}[tbp]
\begin{center}
\caption{\textbf{Quantitative results on RGB panorama generation} compared with state-of-the-art methods, evaluated on both in-domain Sturctured3D and out-of-domain SUN360 datasets.}
\label{tab:comp_rgb}
\resizebox{\linewidth}{!}{
\begin{tabular}{c cc cc c}
    \toprule
    \multirow{2}{*}{{Methods}} & \multicolumn{2}{c}{\textbf{Structured3D}} & \multicolumn{2}{c}{\textbf{SUN360}} & \multirow{2}{*}{\textbf{Avg. Rank}} \\
    & FID $\downarrow$ & IS $\uparrow$ & FID $\downarrow$ & IS $\uparrow$ & \\
    \midrule
    {OmniDreamer$^\dagger$~\cite{omnidreamer}} & 97.46 & 3.35 & 128.17 & 2.29 & 7.0 \\
    {LDM3D-Pano~\cite{ldm3d}} & 32.57 & \underline{6.13} & 72.67 & 4.86 & \underline{3.3} \\
    {Diffusion360$^\dagger$~\cite{diffusion360}} & 26.23 & 4.85 & \textbf{63.03} & 4.21 & 3.5 \\
    {MVDiffusion~\cite{mvdiffusion}} & 35.99 & 5.00 & 67.22 & 4.33 & 4.0\\
    {PanoDiffusion~\cite{panodiffusion}} & \underline{16.20} & 4.04 & 80.02 & 3.91 & 4.8 \\
    {PanFusion~\cite{panfusion}} & 44.86 & \textbf{6.18} & 84.25 & \underline{4.98} & 3.8 \\
    \midrule
    {DreamCube (Ours)} & \textbf{12.58} & 5.50 & \underline{66.56} & \textbf{5.35} & \textbf{1.8} \\
    \bottomrule
\end{tabular}
}
\vspace{-0.3cm}
\footnotesize{$^\dagger$Including SUN360 as training data.}\\
\end{center}
\end{table}

\textbf{Quantitative results for depth panorama generation.} We compare our approach with RGB-D panorama generation methods: LDM3D-Pano~\cite{ldm3d}, PanoDiffusion~\cite{panodiffusion}, and panoramic depth estimation method: Depth Any Camera (DAC)~\cite{depthanycamera}, and the results are reported in Table~\ref{tab:comp_depth}. Our method outperforms the competing methods consistently on all metrics, demonstrating the superiority of the proposed joint RGB-D cube map generation in obtaining accurate geometry compared to equirectangular-based RGB-D generation and depth estimation methods.

\begin{table}[tbp]
\begin{center}
\caption{\textbf{Quantitative results on depth panorama generation} compared with RGB-D panorama generation methods: LDM3D-Pano~\cite{ldm3d}, PanoDiffusion~\cite{panodiffusion}, and panoramic depth estimation method: Depth Any Camera (DAC)~\cite{depthanycamera}.}
\label{tab:comp_depth}
\resizebox{\linewidth}{!}{
\begin{tabular}{c cc  cc}
    \toprule
    Methods & $\delta$-1.25 $\uparrow$ & AbsRel $\downarrow$ & RMSE $\downarrow$ & MAE $\downarrow$ \\
    \midrule
    {LDM3D-Pano~\cite{ldm3d}} & 0.655 & 0.199 & 0.323 & 0.267 \\
    {PanoDiffusion~\cite{panodiffusion}} & 0.695 & 0.160 & 0.301 & 0.255 \\
    {DAC~\cite{depthanycamera}} & 0.751 & 0.139 & 0.266 & 0.220 \\
    \midrule
    {DreamCube (Ours)} & \textbf{0.787} & \textbf{0.133} & \textbf{0.256} & \textbf{0.211} \\
    \bottomrule
\end{tabular}
}
\end{center}
\end{table}

\begin{figure*}[tbp]
\centering
\includegraphics[width=\linewidth]{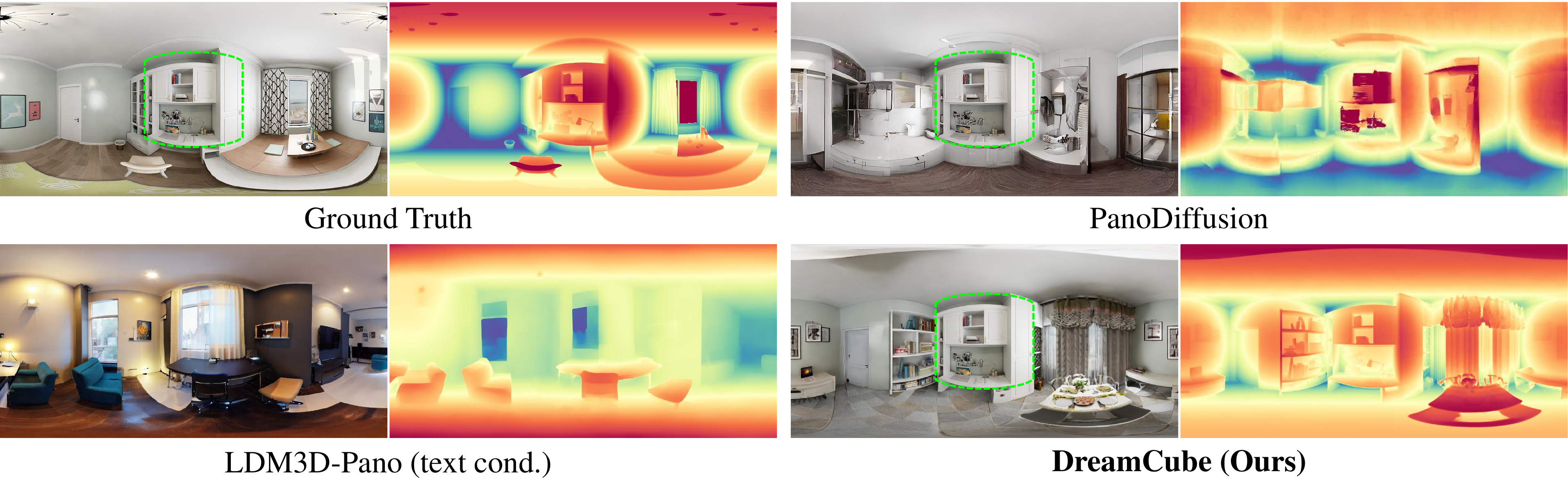}
\caption{\textbf{Qualitative results of the proposed DreamCube} on RGB-D panorama generation compared with RGB-D panorama generation methods: LDM3D-Pano~\cite{ldm3d} and PanoDiffusion~\cite{panodiffusion}. The \textcolor{green}{\textbf{green}} dashed boxes highlight the input image condition.}
\label{fig:comp_rgbd}
\end{figure*}

\textbf{Qualitative results.} We provide visualization results of the generated RGB-D panorama, as shown in Figure~\ref{fig:comp_rgbd}. Compared with the equirectangular based RGB-D panorama generation methods: LDM3D-Pano~\cite{ldm3dvr} and PanoDiffusion~\cite{panodiffusion}, our method is able to obtain more detailed and accurate geometry. Moreover, our generated RGB panoramas have fewer artifacts than PanoDiffusion which uses the same training set as ours. We attribute these advantages to the fact that the cube map representations can better exploit the 2D diffusion's image priors compared to the geometrically distorted equirectangular representations.

\begin{figure*}[tbp]
\centering
\includegraphics[width=\linewidth]{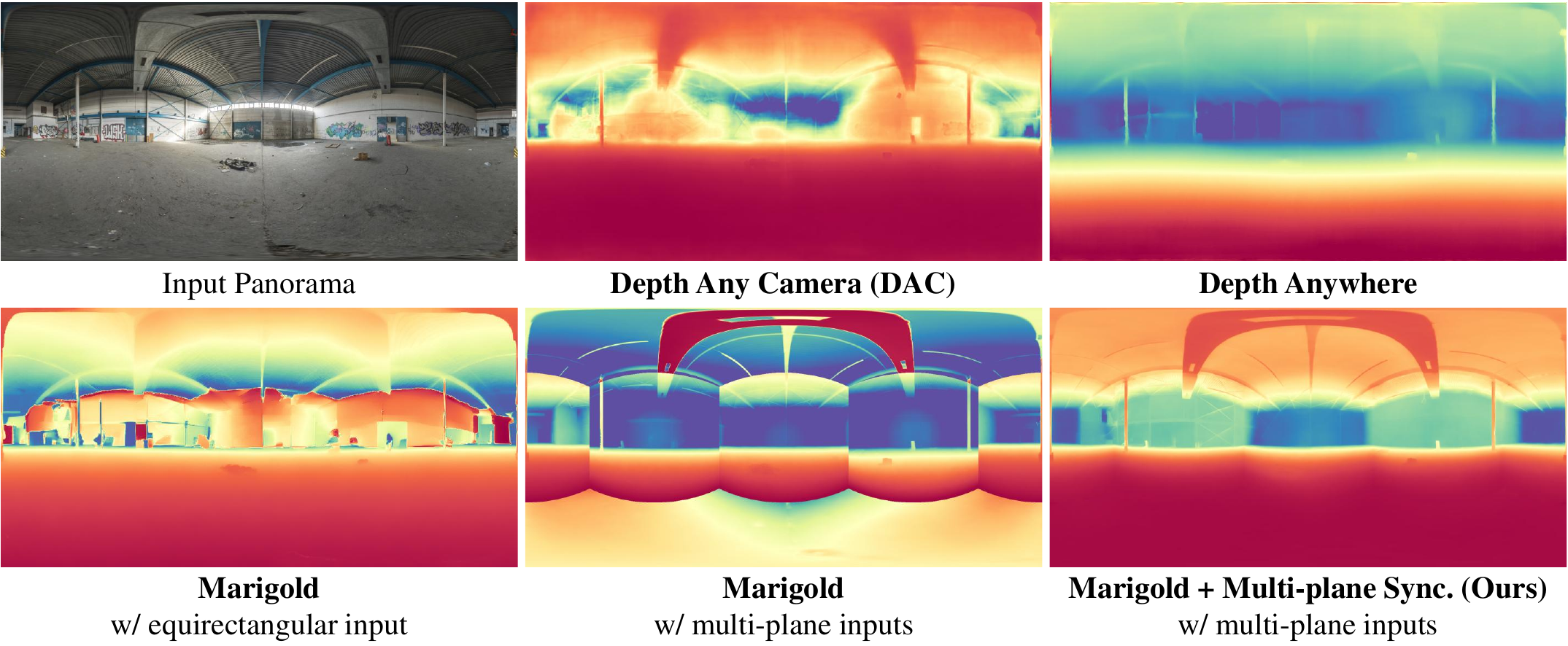}
\caption{\textbf{Qualitative results of the proposed Multi-plane Synchronization} on panoramic depth estimation compared with recent panoramic depth estimation methods: Depth Any Camera (DAC)~\cite{depthanycamera} and Depth Anywhere~\cite{depthanywhere}.}
\label{fig:sync_depth}
\end{figure*}

\begin{figure*}[tbp]
\centering
\includegraphics[width=\linewidth]{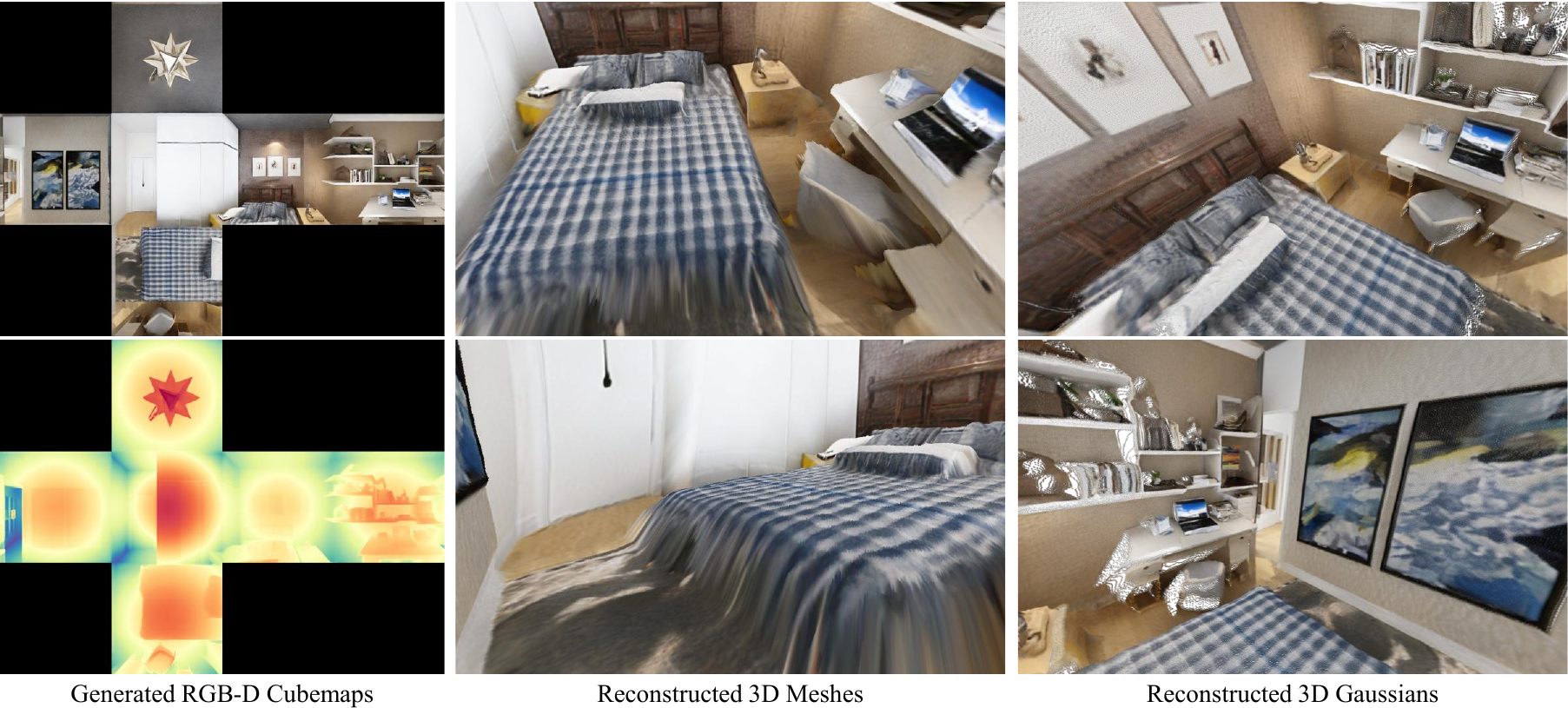}
\caption{\textbf{Panorama-to-3D scene reconstruction.} Based on the RGB-D cubemap generated by DreamCube, we can reconstruct the corresponding 3D scenes in seconds and obtain both 3D mesh and 3D Gaussian~\cite{3dgs} representations.}
\label{fig:pano2scene}
\end{figure*}

\begin{figure}[tbp]
\centering
\includegraphics[width=\linewidth]{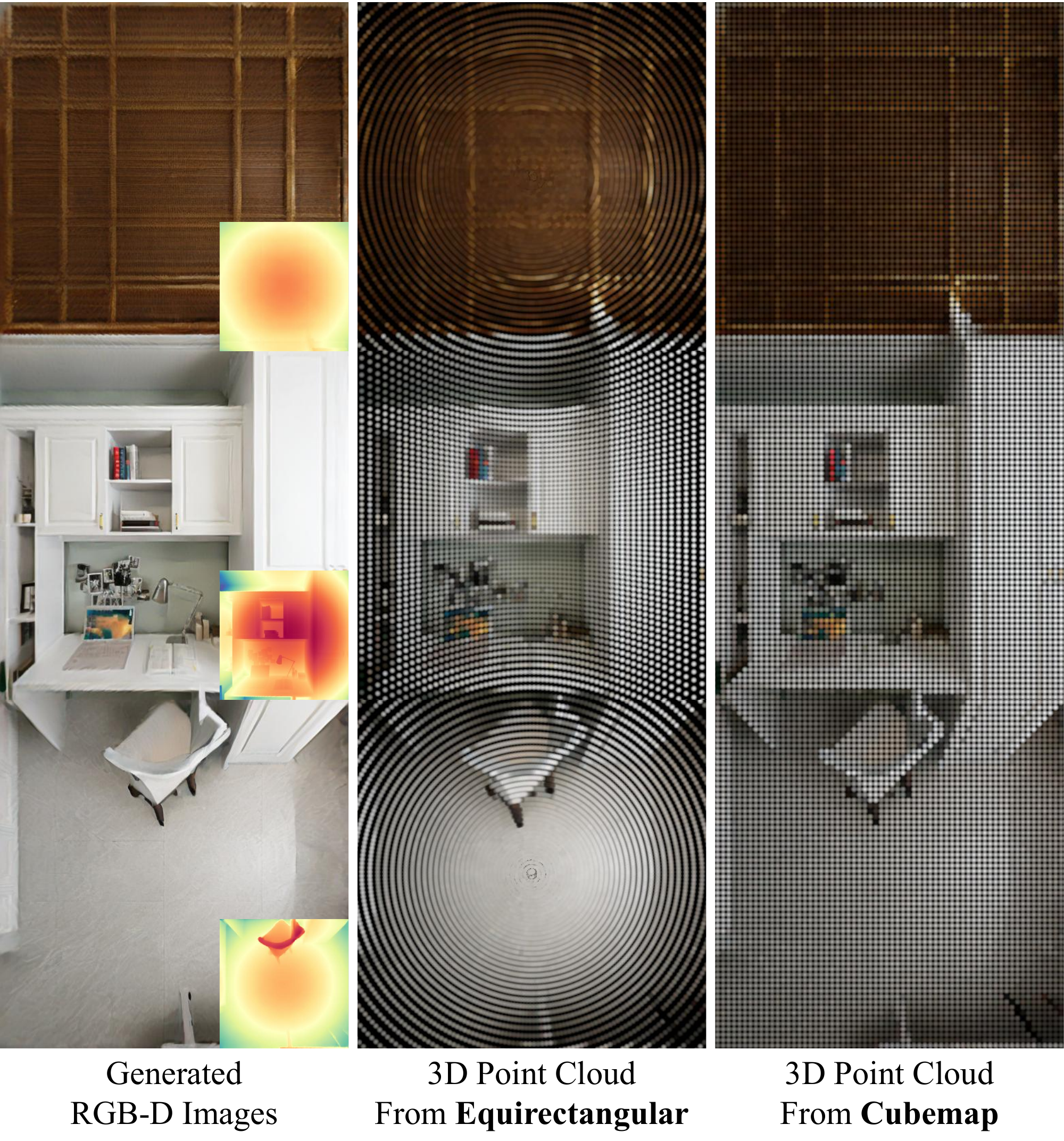}
\caption{\textbf{Qualitative comparison of 3D point clouds} reconstructed from equirectangular-based and cubemap-based RGB-D panoramas. Equirectangular panoramas produce an uneven, ring-shaped 3D point distribution dense near the poles, while cubemap panoramas yield a more uniform and regular distribution.}
\label{fig:pano2scene_comp}
\end{figure}

\subsection{Panoramic Depth Estimation}
Our proposed Multi-plane Synchronization extends naturally to monocular depth estimation, demonstrating its versatility beyond RGB panorama generation.
As shown in Figure~\ref{fig:sync_depth}, applying our synchronization approach to Marigold effectively eliminates the discontinuous seams that occur when processing multi-plane inputs with the original model.
Qualitative comparisons with recent panoramic depth estimation methods: Depth Any Camera (DAC)~\cite{depthanycamera} and Depth Anywhere~\cite{depthanywhere} reveal that our approach captures more detailed geometric structures while maintaining the continuity of depth values at left-right seams (which Depth Anywhere fails).
These results show that the proposed Multi-plane Synchronization can effectively generalize diffusion-based monocular depth estimation models to multi-plane omnidirectional representations with painless modification and minimal performance loss.

\subsection{Panorama-to-3D Reconstruction}
An important application of DreamCube is fast 3D scene generation. Benefiting from the joint RGB-D panorama generation model and the rapid panorama-to-3D projection algorithm, our approach can achieve 3D scene generation from a single view in about ten seconds. We present the visualized results of the generated 3D scenes in both 3D meshes and 3D Gaussian representations, as shown in Figure~\ref{fig:pano2scene}. The visual quality of the reconstructed 3D scene is comparable to that of the original panorama. Additionally, we analyze the impact of different formats of RGB-D panoramas on 3D reconstruction, as illustrated in Figure~\ref{fig:pano2scene_comp}. The 3D point distribution derived from equirectangular-based panoramas is uneven, exhibiting a ring-shaped pattern with particularly dense points near the poles. In contrast, the 3D point distribution from cubemap-based panoramas tends to be more uniform and regular.

\subsection{Ablation Study and Analysis}
We perform a series of analyses on the proposed {Multi-plane Synchronization} and {DreamCube} to evaluate their effectiveness and robustness under various conditions.

\begin{figure*}[tbp]
\centering
\includegraphics[width=\linewidth]{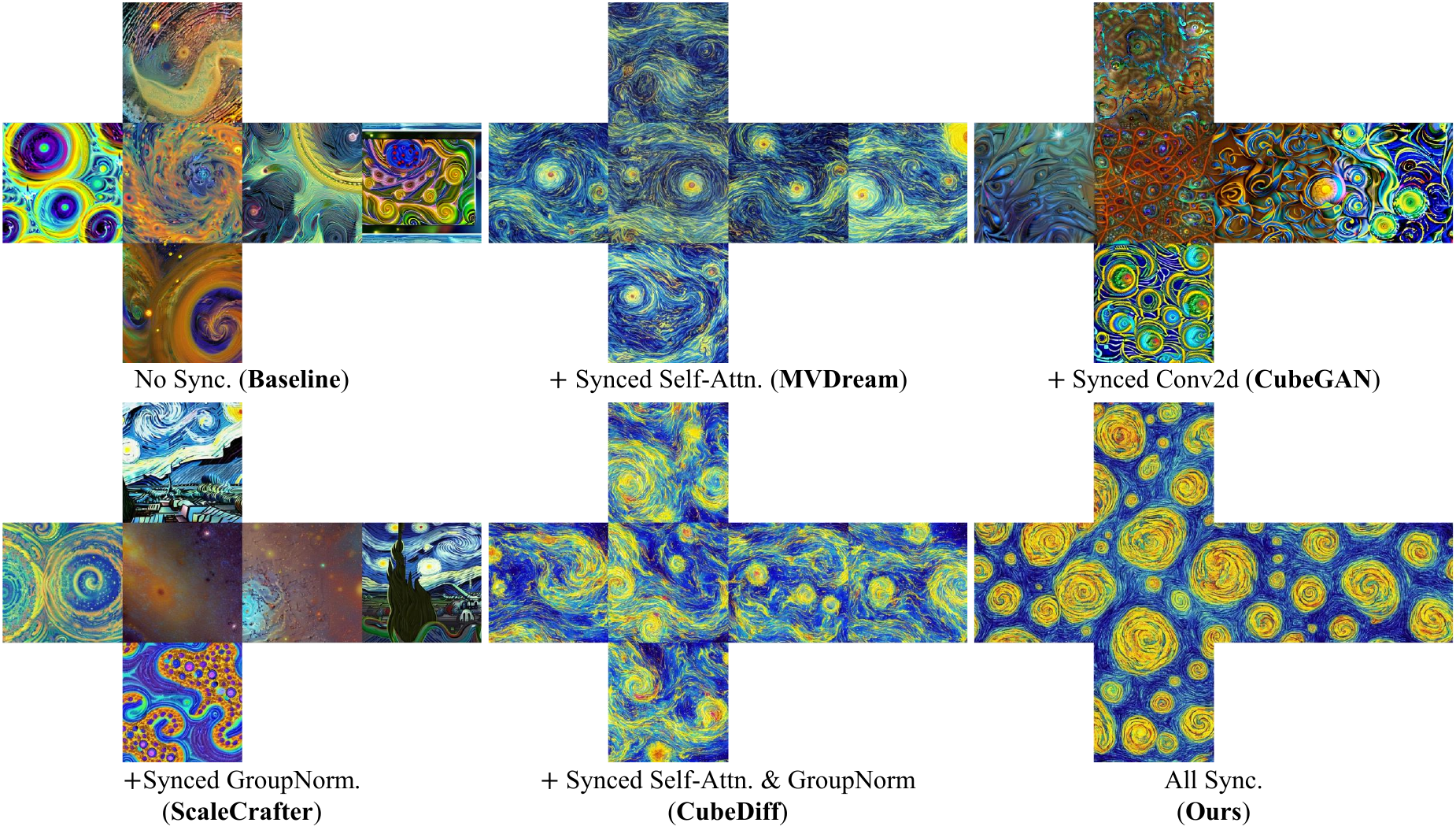}
\caption{\textbf{Ablation analysis of Multi-plane Synchronization.} We adopt Stable-Diffusion v2 as baseline model for multi-plane generation, and the text prompt used is ``The vast cosmos in the style of Van Gogh, with swirling patterns and vibrant colors.''}
\label{fig:sync_ablation}
\end{figure*}

\textbf{Ablation analysis of Multi-plane Synchronization.} 
To analyze the impact of different synced operators, we perform different synchronization configurations and provide the visualization results in Figure~\ref{fig:sync_ablation}. 
Synced self-attention ensures content consistency across faces, synced group normalization harmonizes color tones, and synced convolutions reduce seam discontinuities.
Combining all three operators produces panoramas with seamless transitions, consistent content, and uniform style.

\begin{table}[tbp]
    \begin{center}
    \caption{\textbf{Ablation analysis of DreamCube}, where the performance evaluation is performed on the Structured3D test split~\cite{structured3d}. Both proposed Multi-plane Synchronization and XYZ Positional encoding bring performance improvements.}
    \label{tab:ablation}
    \resizebox{\linewidth}{!}{
    \begin{tabular}{c cc  cc}
        \toprule
        \multirow{2}{*}{{Methods}} & \multicolumn{2}{c}{{RGB}} & \multicolumn{2}{c}{{Depth}}\\
         & FID $\downarrow$ & IS $\uparrow$ & $\delta$-1.25 $\uparrow$ & AbsRel $\downarrow$\\
        \midrule
        w/o XYZ Pos. & 13.66 & 5.57 & 0.784 & 0.136 \\
        w/o Sync. & 21.35 & {5.62} & 0.684 & 0.168 \\
        \midrule
        {w/o SyncSA} &24.38 &\textbf{5.78} &0.715 &0.158 \\
        {w/o SyncConv} &19.62 &5.60 & 0.779 & 0.139 \\
        {w/o SyncGN} &18.35 &5.51 & 0.784 & 0.135 \\
        \midrule
        DreamCube (Ours) & \textbf{12.58} & {5.50} & \textbf{0.787} & \textbf{0.133} \\
        \bottomrule
    \end{tabular}
    }
    \end{center}
\end{table}


\begin{figure}[tbp]
\centering
\includegraphics[width=\linewidth]{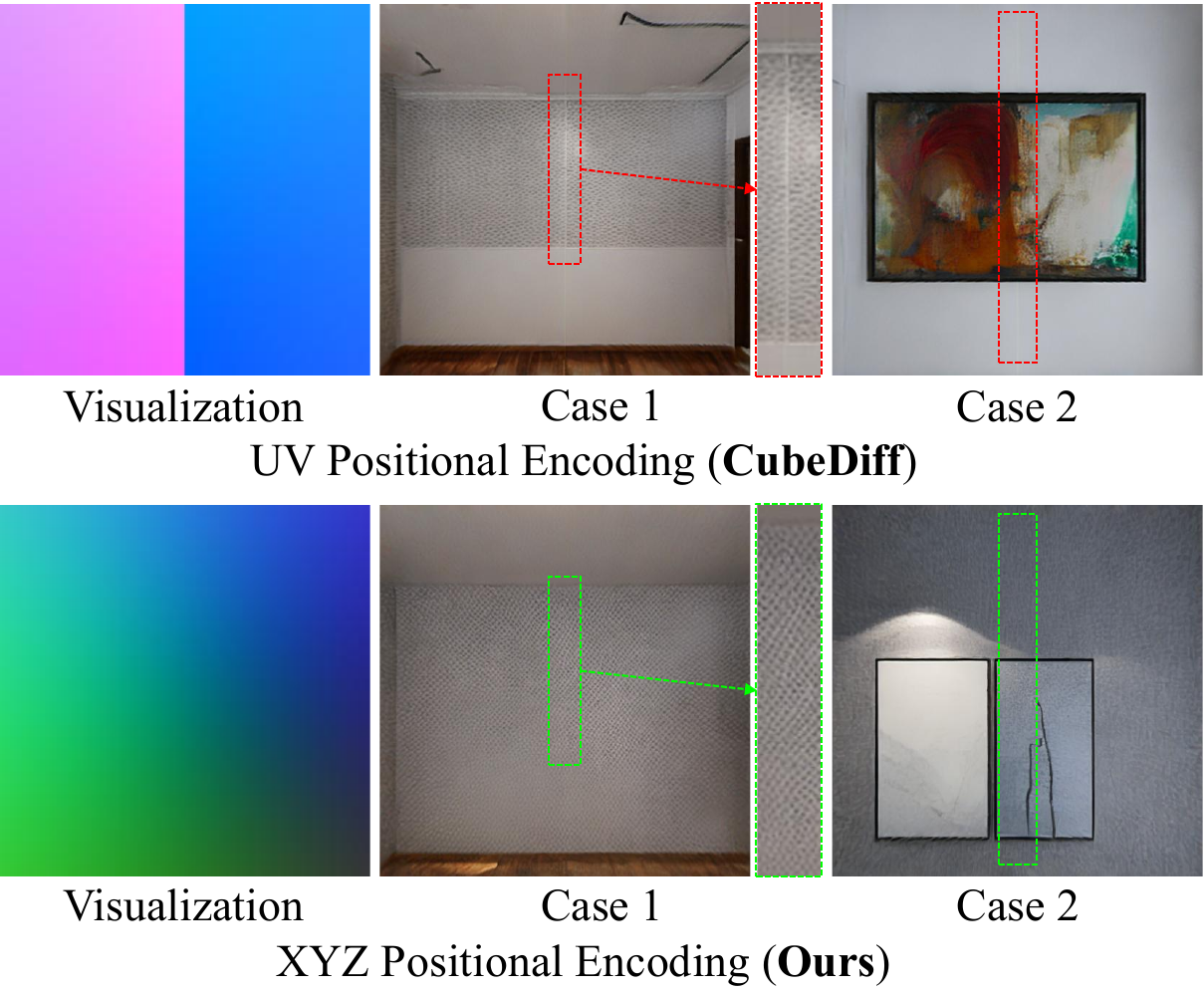}
\caption{\textbf{Ablation analysis of XYZ Positional Encoding.} We present the qualitative results of the back view of cubemap, where the UV positional encoding introduces discontinuous numerical steps. This leads to line artifacts (Case 1) and incoherent visual contents (Case 2), as indicated by the \textcolor{red}{\textbf{red}} dashed box. In contrast, our proposed XYZ positional encoding alleviates these issues in both cases, as shown within the \textcolor{green}{\textbf{green}} dashed box.}
\label{fig:pos_encoding_ablation}
\end{figure}

\textbf{Ablation analysis of DreamCube.} 
We analyze different components of DreamCube, with results shown in Table~\ref{tab:ablation}.
We evaluate both RGB and depth panorama generation on the Structured3D test split~\cite{structured3d}, following evaluation protocol in Sec.~\ref{sec:eval_rgbd}.
Both XYZ Positional Encoding (``XYZ Pos.'') and Multi-plane Synchronization (``Sync.'') improve performance, with ``Sync.'' yielding the most substantial gains, which reduces FID by 8.77 and improves $\delta$-1.25 by 0.103. Specifically, Synced Self-Attention (``SyncSA'') contributes the most performance gain compared to other synced operators. Besides, we further provide a qualitative ablation analysis of XYZ Positional Encoding in Figure~\ref{fig:pos_encoding_ablation}. Our design effectively alleviates line artifacts and content incoherence compared to UV Positional Encoding.

\begin{figure*}[tbp]
\centering
\includegraphics[width=\linewidth]{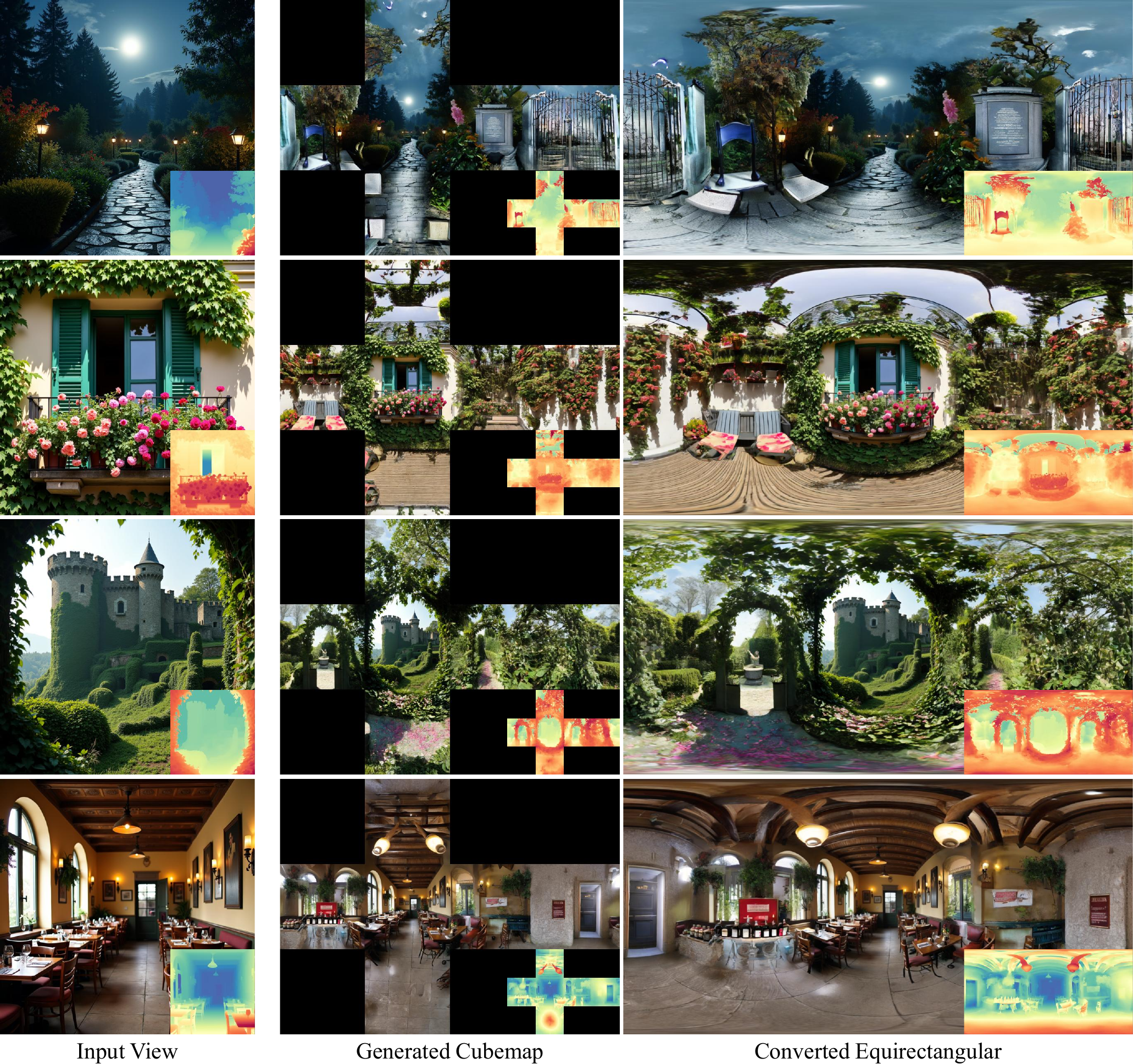}
\caption{\textbf{Out-domain RGB-D panorama generation.} The RGB-D inputs are obtained by Flux.1-dev~\cite{flux1dev} and Depth Anything v2~\cite{depthanythingv2}. DreamCube demonstrates generalization ability on diverse inputs, maintaining high-quality RGB appearance and geometric consistency.}
\label{fig:more_demo}
\end{figure*}

\textbf{Generalization analysis of DreamCube.} To evaluate DreamCube's generalization capabilities, we present out-domain generation results in Figure~\ref{fig:more_demo}, where the input RGB images are generated from Flux.1-dev~\cite{flux1dev}. We obtain the corresponding input depth map using Depth Anything v2~\cite{depthanythingv2}. Despite the significant domain gap between these inputs and our training distribution, DreamCube successfully generates coherent and visually plausible RGB-D panoramas, demonstrating its strong generalization ability.

\begin{figure}[tbp]
  \centering
  \begin{subfigure}{\linewidth}
    \centering
    \includegraphics[width=\linewidth]{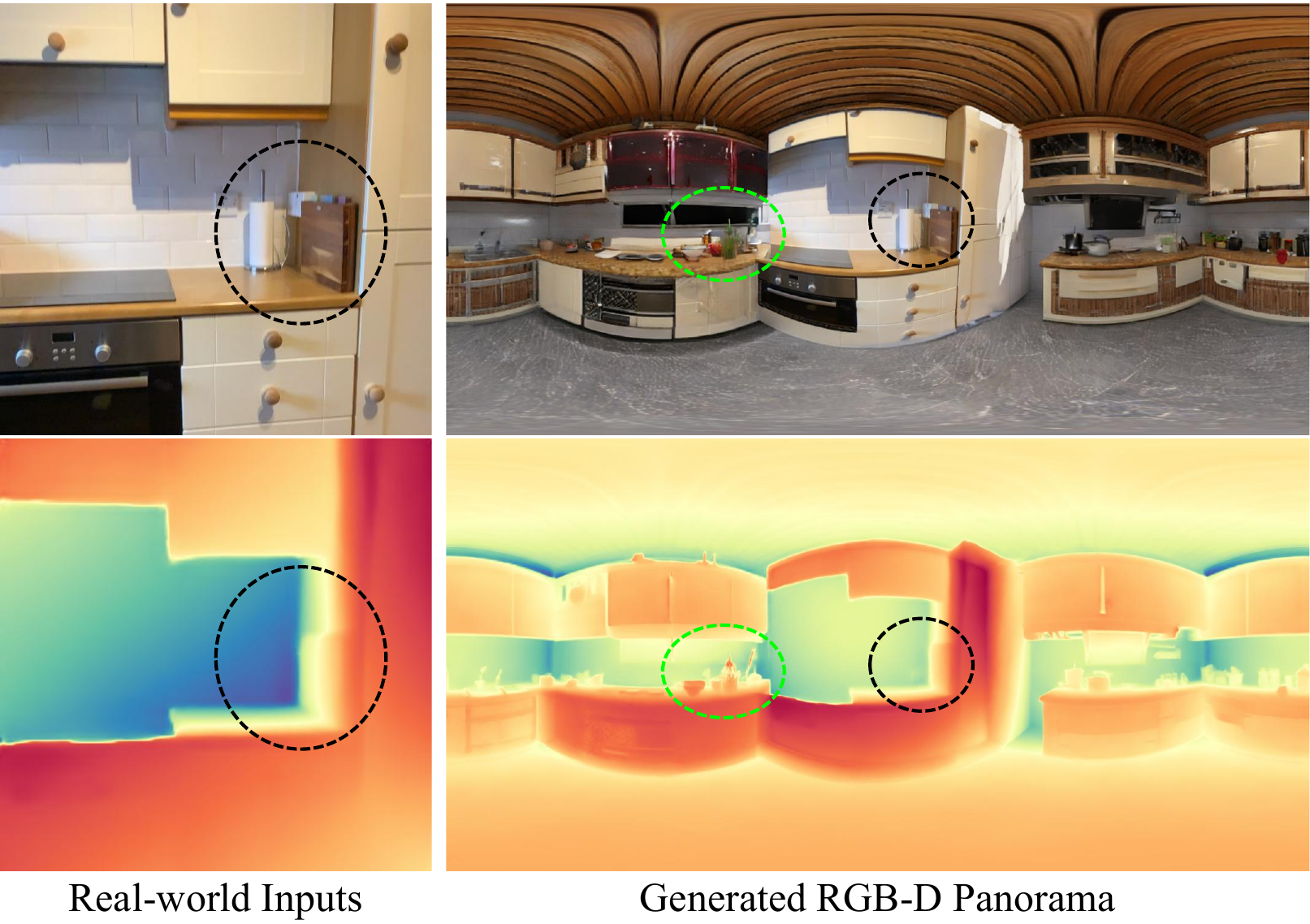}
    \caption{Generated results from real-world inputs captured by sensors~\cite{arkitscenes}. Even though the input depth is low-resolution (as indicated by the \textbf{black} dashed circles), our method is still able to generate high-definition depth maps (as indicated by the \textcolor{green}{\textbf{green}} dashed circles).}
    \label{fig:robustness_input_real_world}
    \vspace{4pt}
  \end{subfigure}
  \begin{subfigure}{\linewidth}
    \centering
    \includegraphics[width=\linewidth]{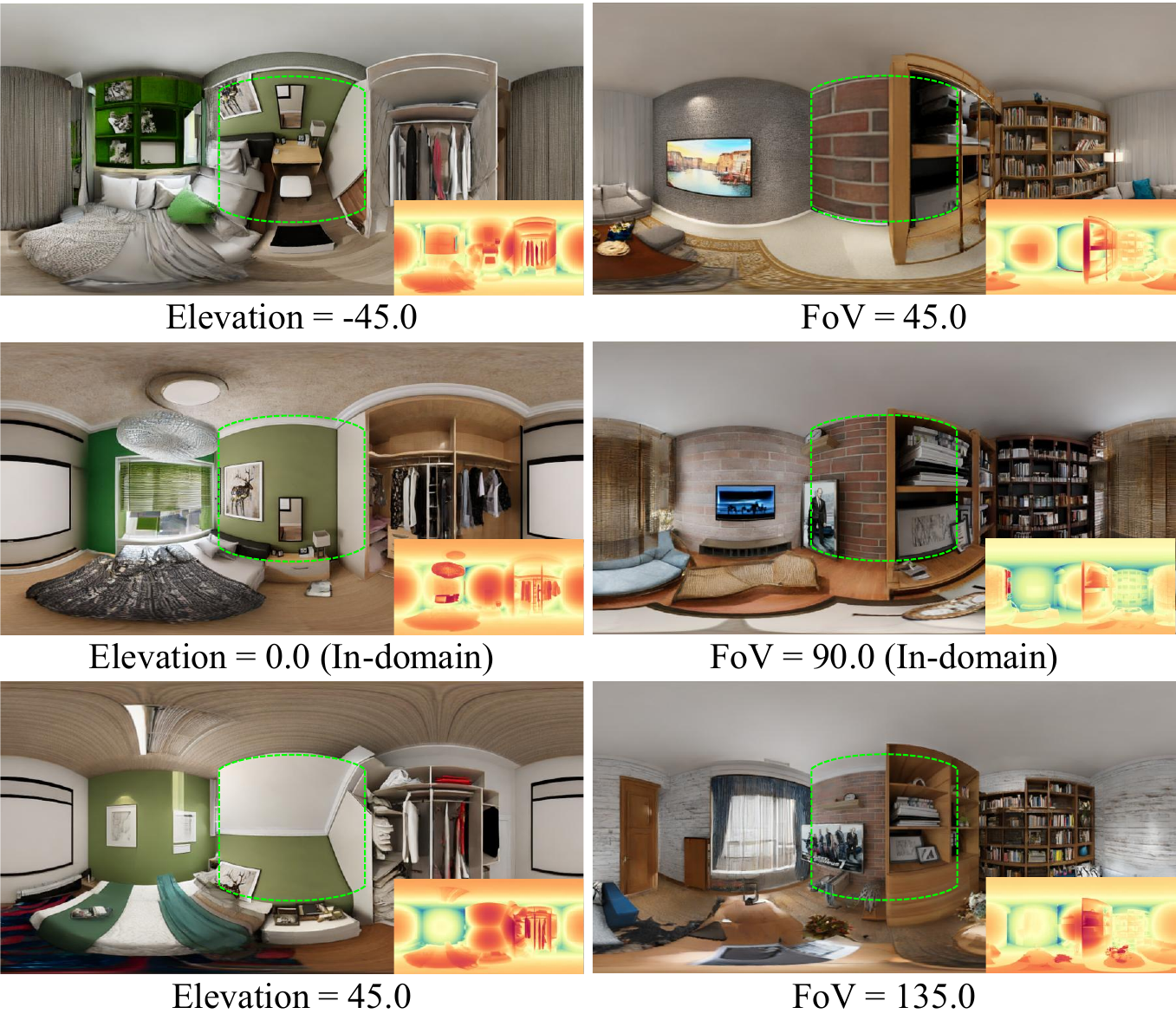}
    \caption{Generated results from inputs with extreme viewing angles, where the \textcolor{green}{\textbf{green}} dashed boxes highlight the input views.}
    \label{fig:robustness_input_view}
  \end{subfigure}
  \caption{\textbf{Robustness analysis} of DreamCube to out-domain RGB-D inputs from real world and extreme viewing angles. }
  \label{fig:robustness_input}
\end{figure}

\textbf{Robustness analysis of DreamCube.} DreamCube takes single-view RGB-D images as input for cubemap generation. To evaluate the robustness of DreamCube, we test various types of RGB-D inputs and provide the generated results in Figure~\ref{fig:robustness_input}. Specifically, we test real-world inputs captured by sensors~\cite{arkitscenes}. Unlike synthetic training data, real-world inputs have low-resolution depth maps and non-standard camera views. Even so, our method is still able to generate reasonable panoramas with high-resolution depth maps, as shown in Figure~\ref{fig:robustness_input_real_world}. In addition, we also test inputs with extreme camera views (\eg, elevation and FoV). DreamCube struggles to generate correct panoramas under inputs with extreme elevation angles, but shows robustness to perturbations of the FoV, as shown in Figure~\ref{fig:robustness_input_view}.

\textbf{Efficiency analysis.} We provide an efficiency analysis of our approach in Table~\ref{tab:efficiency} compared to the baseline model, Stable Diffusion v2 (SD2)~\cite{latentdiffusion}. Among all synchronized operators, synchronized Self-Attention (``{+SyncSA}'') incurs the most computational cost, increasing TFLOPs by 76.1\% and latency (ms) by 113.1\% than no synchronization (``{No Sync.}''). This accounts for 86.0\% of the latency cost and almost 100\% of the TFLOPs cost incurred by our approach.

\begin{table}[tbp]
\begin{center}
\caption{\textbf{Efficiency analysis.} We evaluate the computational efficiency of SD2’s and DreamCube’s U-Nets in a single forward pass and report the metrics: TFLOPs and Latency (ms).}
\label{tab:efficiency}
\resizebox{\linewidth}{!}{
\begin{tabular}{ cc cc}
\toprule
\multicolumn{2}{c}{Methods} & {FLOPs (T)} & Latency (ms) \\
\midrule
\multirow{2}{*}{SD2's U-Net} &batch-size=1 &0.804 &35.4 \\
&batch-size=6 &4.826 &138.8 \\
\midrule
\multirow{5}{*}{\makecell{DreamCube's\\U-Net}} &{No Sync.} &4.827 &139.4 \\
 &{+SyncSA} &8.502 &297.1 \\
 &{+SyncConv} &4.827 &144.3 \\
 &{+SyncGN} &4.827 &152.0 \\
 &{All Sync.} &8.502 &322.7 \\
\bottomrule
\end{tabular}
}
\end{center}
\end{table}

%% file: sections/5_conclusion.tex
\section{Limitation}

The limitations of our method include high computational cost and the restricted input conditions.
First, DreamCube samples six image latents at a time, which incurs additional computational cost and hinders the scalability of training batches. Nevertheless, compared with existing panorama generation methods, our method has the superior computational utilization (effective pixel ratio obtained at the same computational cost), because it uses a less distorted cubemap instead of equirectangular, and does not require redundant FoV overlapping for seam continuity.
Second, DreamCube is trained with cubemap's front face as input conditions. When the input distribution deviates from the training domain, for example, non-frontal view or extreme FoV, the generated results may fail, as shown in Figure~\ref{fig:robustness_input}.

\section{Conclusion}

In this work, we thoroughly analyze the limitations of existing 2D diffusion models when applied to multi-plane panoramic representations, and propose a multi-plane synchronization strategy to painlessly generalize pre-trained 2D diffusion models to the omnidirectional image domain. This strategy ensures translation equivariance in the omnidirectional image domain by adapting the spatial operators in the model without the need for additional fine-tuning or constructing overlapping views. Benefiting from multi-plane synchronization, we further introduce DreamCube, which jointly models RGB and depth cube maps by leveraging image priors from pre-trained 2D diffusion. Extensive experiments show that the proposed approach outperforms previous equirectangular-based RGB-D panoramic generation methods and holds potential in panoramic depth estimation and 3D scene generation.

%% file: main.bbl
\begin{thebibliography}{65}
\providecommand{\natexlab}[1]{#1}
\providecommand{\url}[1]{\texttt{#1}}
\expandafter\ifx\csname urlstyle\endcsname\relax
  \providecommand{\doi}[1]{doi: #1}\else
  \providecommand{\doi}{doi: \begingroup \urlstyle{rm}\Url}\fi

\bibitem[Akimoto et~al.(2022)Akimoto, Matsuo, and Aoki]{omnidreamer}
Naofumi Akimoto, Yuhi Matsuo, and Yoshimitsu Aoki.
\newblock {Diverse Plausible 360-Degree Image Outpainting for Efficient 3DCG Background Creation}.
\newblock In \emph{Proceedings of the IEEE/CVF Conference on Computer Vision and Pattern Recognition (CVPR)}, 2022.

\bibitem[Bar-Tal et~al.(2023)Bar-Tal, Yariv, Lipman, and Dekel]{multidiffusion}
Omer Bar-Tal, Lior Yariv, Yaron Lipman, and Tali Dekel.
\newblock {MultiDiffusion: Fusing Diffusion Paths for Controlled Image Generation}.
\newblock In \emph{International Conference on Machine Learning}, pages 1737--1752. PMLR, 2023.

\bibitem[Baruch et~al.(2021)Baruch, Chen, Dehghan, Dimry, Feigin, Fu, Gebauer, Joffe, Kurz, Schwartz, and Shulman]{arkitscenes}
Gilad Baruch, Zhuoyuan Chen, Afshin Dehghan, Tal Dimry, Yuri Feigin, Peter Fu, Thomas Gebauer, Brandon Joffe, Daniel Kurz, Arik Schwartz, and Elad Shulman.
\newblock {{ARK}itScenes - A Diverse Real-World Dataset for 3D Indoor Scene Understanding Using Mobile {RGB}-D Data}.
\newblock In \emph{Neural Information Processing Systems}, 2021.

\bibitem[Chen et~al.(2022)Chen, Wang, and Liu]{text2light}
Zhaoxi Chen, Guangcong Wang, and Ziwei Liu.
\newblock Text2light: Zero-shot text-driven hdr panorama generation.
\newblock \emph{ACM Transactions on Graphics (TOG)}, 41\penalty0 (6):\penalty0 1--16, 2022.

\bibitem[Cheng et~al.(2018)Cheng, Wang, and Yang]{cheng2018depth}
Xinjing Cheng, Peng Wang, and Ruigang Yang.
\newblock Depth estimation via affinity learned with convolutional spatial propagation network.
\newblock In \emph{Proceedings of the European conference on computer vision (ECCV)}, pages 103--119, 2018.

\bibitem[Duan et~al.(2024)Duan, Guo, and Zhu]{diffusiondepth}
Yiquan Duan, Xianda Guo, and Zheng Zhu.
\newblock Diffusiondepth: Diffusion denoising approach for monocular depth estimation.
\newblock In \emph{European Conference on Computer Vision}, pages 432--449. Springer, 2024.

\bibitem[Eigen and Fergus(2015)]{eigen2015predicting}
David Eigen and Rob Fergus.
\newblock Predicting depth, surface normals and semantic labels with a common multi-scale convolutional architecture.
\newblock In \emph{Proceedings of the IEEE international conference on computer vision}, pages 2650--2658, 2015.

\bibitem[Esser et~al.(2024)Esser, Kulal, Blattmann, Entezari, M{\"u}ller, Saini, Levi, Lorenz, Sauer, Boesel, et~al.]{esser2024scaling}
Patrick Esser, Sumith Kulal, Andreas Blattmann, Rahim Entezari, Jonas M{\"u}ller, Harry Saini, Yam Levi, Dominik Lorenz, Axel Sauer, Frederic Boesel, et~al.
\newblock Scaling rectified flow transformers for high-resolution image synthesis.
\newblock In \emph{Forty-first international conference on machine learning}, 2024.

\bibitem[Feng et~al.(2023)Feng, Liu, Cui, and Xie]{diffusion360}
Mengyang Feng, Jinlin Liu, Miaomiao Cui, and Xuansong Xie.
\newblock {Diffusion360: Seamless 360 Degree Panoramic Image Generation based on Diffusion Models}.
\newblock \emph{arXiv preprint arXiv:2311.13141}, 2023.

\bibitem[Fu et~al.(2024)Fu, Yin, Hu, Wang, Ma, Tan, Shen, Lin, and Long]{geowizard}
Xiao Fu, Wei Yin, Mu Hu, Kaixuan Wang, Yuexin Ma, Ping Tan, Shaojie Shen, Dahua Lin, and Xiaoxiao Long.
\newblock Geowizard: Unleashing the diffusion priors for 3d geometry estimation from a single image.
\newblock In \emph{European Conference on Computer Vision}, pages 241--258. Springer, 2024.

\bibitem[Goodfellow et~al.(2020)Goodfellow, Pouget-Abadie, Mirza, Xu, Warde-Farley, Ozair, Courville, and Bengio]{gan}
Ian Goodfellow, Jean Pouget-Abadie, Mehdi Mirza, Bing Xu, David Warde-Farley, Sherjil Ozair, Aaron Courville, and Yoshua Bengio.
\newblock Generative adversarial networks.
\newblock \emph{Communications of the ACM}, 63\penalty0 (11):\penalty0 139--144, 2020.

\bibitem[Guo et~al.(2025)Guo, Garg, Miangoleh, Huang, and Ren]{depthanycamera}
Yuliang Guo, Sparsh Garg, S~Mahdi~H Miangoleh, Xinyu Huang, and Liu Ren.
\newblock Depth any camera: Zero-shot metric depth estimation from any camera.
\newblock \emph{arXiv preprint arXiv:2501.02464}, 2025.

\bibitem[hdri skies(accessed 02/2025{\natexlab{a}})]{hdriskies}
hdri skies.
\newblock {HDRIs}.
\newblock \emph{https://hdri-skies.com/}, accessed 02/2025{\natexlab{a}}.

\bibitem[hdri skies(accessed 02/2025{\natexlab{b}})]{ihdri}
hdri skies.
\newblock {HDRIs}.
\newblock \emph{https://www.ihdri.com/hdri-skies-outdoor/}, accessed 02/2025{\natexlab{b}}.

\bibitem[He et~al.(2024{\natexlab{a}})He, Li, Yin, Liang, Li, Zhou, Zhang, Liu, and Chen]{lotus}
Jing He, Haodong Li, Wei Yin, Yixun Liang, Leheng Li, Kaiqiang Zhou, Hongbo Zhang, Bingbing Liu, and Ying-Cong Chen.
\newblock Lotus: Diffusion-based visual foundation model for high-quality dense prediction.
\newblock \emph{arXiv preprint arXiv:2409.18124}, 2024{\natexlab{a}}.

\bibitem[He et~al.(2024{\natexlab{b}})He, Yang, Chen, Cun, Xia, Zhang, Wang, He, Chen, and Shan]{scalecrafter}
Yingqing He, Shaoshu Yang, Haoxin Chen, Xiaodong Cun, Menghan Xia, Yong Zhang, Xintao Wang, Ran He, Qifeng Chen, and Ying Shan.
\newblock {ScaleCrafter: Tuning-free Higher-Resolution Visual Generation with Diffusion Models}.
\newblock In \emph{International Conference on Learning Representations}, 2024{\natexlab{b}}.

\bibitem[Heusel et~al.(2017)Heusel, Ramsauer, Unterthiner, Nessler, and Hochreiter]{heusel2017gans}
Martin Heusel, Hubert Ramsauer, Thomas Unterthiner, Bernhard Nessler, and Sepp Hochreiter.
\newblock Gans trained by a two time-scale update rule converge to a local nash equilibrium.
\newblock \emph{Advances in neural information processing systems}, 30, 2017.

\bibitem[Ho et~al.(2020)Ho, Jain, and Abbeel]{ddpm}
Jonathan Ho, Ajay Jain, and Pieter Abbeel.
\newblock {Denoising Diffusion Probabilistic Models}.
\newblock \emph{Advances in Neural Information Processing Systems}, 33:\penalty0 6840--6851, 2020.

\bibitem[Ji et~al.(2025)Ji, Zoss, Chandran, Yang, Cao, Solenthaler, and Bradley]{ji2025joint}
Xinya Ji, Gaspard Zoss, Prashanth Chandran, Lingchen Yang, Xun Cao, Barbara Solenthaler, and Derek Bradley.
\newblock Joint learning of depth and appearance for portrait image animation.
\newblock \emph{arXiv preprint arXiv:2501.08649}, 2025.

\bibitem[Ji et~al.(2023)Ji, Chen, Xie, Hong, Liu, Liu, Lu, Li, and Luo]{ddp}
Yuanfeng Ji, Zhe Chen, Enze Xie, Lanqing Hong, Xihui Liu, Zhaoqiang Liu, Tong Lu, Zhenguo Li, and Ping Luo.
\newblock Ddp: Diffusion model for dense visual prediction.
\newblock In \emph{Proceedings of the IEEE/CVF International Conference on Computer Vision}, pages 21741--21752, 2023.

\bibitem[Kalischek et~al.(2025)Kalischek, Oechsle, Manhardt, Henzler, Schindler, and Tombari]{cubediff}
Nikolai Kalischek, Michael Oechsle, Fabian Manhardt, Philipp Henzler, Konrad Schindler, and Federico Tombari.
\newblock Cubediff: Repurposing diffusion-based image models for panorama generation, 2025.

\bibitem[Ke et~al.(2024)Ke, Obukhov, Huang, Metzger, Daudt, and Schindler]{marigold}
Bingxin Ke, Anton Obukhov, Shengyu Huang, Nando Metzger, Rodrigo~Caye Daudt, and Konrad Schindler.
\newblock Repurposing diffusion-based image generators for monocular depth estimation.
\newblock In \emph{Proceedings of the IEEE/CVF Conference on Computer Vision and Pattern Recognition (CVPR)}, 2024.

\bibitem[Kerbl et~al.(2023)Kerbl, Kopanas, Leimk{\"u}hler, and Drettakis]{3dgs}
Bernhard Kerbl, Georgios Kopanas, Thomas Leimk{\"u}hler, and George Drettakis.
\newblock {3D Gaussian Splatting for Real-Time Radiance Field Rendering}.
\newblock \emph{ACM Transactions on Graphics}, 42\penalty0 (4), 2023.

\bibitem[Khirodkar et~al.(2024)Khirodkar, Bagautdinov, Martinez, Zhaoen, James, Selednik, Anderson, and Saito]{sapiens}
Rawal Khirodkar, Timur Bagautdinov, Julieta Martinez, Su Zhaoen, Austin James, Peter Selednik, Stuart Anderson, and Shunsuke Saito.
\newblock Sapiens: Foundation for human vision models.
\newblock \emph{arXiv preprint arXiv:2408.12569}, 2024.

\bibitem[Kocabas et~al.(2021)Kocabas, Huang, Tesch, M{\"u}ller, Hilliges, and Black]{pano360}
Muhammed Kocabas, Chun-Hao~P Huang, Joachim Tesch, Lea M{\"u}ller, Otmar Hilliges, and Michael~J Black.
\newblock {SPEC: Seeing People in the Wild With an Estimated Camera}.
\newblock In \emph{Proceedings of the IEEE/CVF International Conference on Computer Vision}, pages 11035--11045, 2021.

\bibitem[Krishnan et~al.(2025)Krishnan, Yan, Casser, and Kundu]{orchid}
Akshay Krishnan, Xinchen Yan, Vincent Casser, and Abhijit Kundu.
\newblock Orchid: Image latent diffusion for joint appearance and geometry generation.
\newblock \emph{arXiv preprint arXiv:2501.13087}, 2025.

\bibitem[Labs(2025{\natexlab{a}})]{flux1dev}
Black~Forest Labs.
\newblock Flux.1-dev.
\newblock \url{https://huggingface.co/black-forest-labs/FLUX.1-dev}, 2025{\natexlab{a}}.
\newblock Accessed: 2025-01-19.

\bibitem[Labs(2025{\natexlab{b}})]{flux2024}
Black~Forest Labs.
\newblock Flux.
\newblock \url{https://github.com/black-forest-labs/flux}, 2025{\natexlab{b}}.
\newblock Accessed: 2025-01-19.

\bibitem[Li et~al.(2015)Li, Shen, Dai, Van Den~Hengel, and He]{li2015depth}
Bo Li, Chunhua Shen, Yuchao Dai, Anton Van Den~Hengel, and Mingyi He.
\newblock Depth and surface normal estimation from monocular images using regression on deep features and hierarchical crfs.
\newblock In \emph{Proceedings of the IEEE conference on computer vision and pattern recognition}, pages 1119--1127, 2015.

\bibitem[Li and Bansal(2023)]{panogen}
Jialu Li and Mohit Bansal.
\newblock {PanoGen: Text-Conditioned Panoramic Environment Generation for Vision-and-Language Navigation}.
\newblock In \emph{International Conference on Neural Information Processing Systems}, 2023.

\bibitem[Li et~al.(2023)Li, Li, Savarese, and Hoi]{blip2}
Junnan Li, Dongxu Li, Silvio Savarese, and Steven Hoi.
\newblock Blip-2: Bootstrapping language-image pre-training with frozen image encoders and large language models.
\newblock In \emph{International conference on machine learning}, pages 19730--19742. PMLR, 2023.

\bibitem[Lin et~al.(2024)Lin, Peng, Chen, Peng, Sun, Liu, Bao, Feng, Zhou, and Kang]{promptda}
Haotong Lin, Sida Peng, Jingxiao Chen, Songyou Peng, Jiaming Sun, Minghuan Liu, Hujun Bao, Jiashi Feng, Xiaowei Zhou, and Bingyi Kang.
\newblock Prompting depth anything for 4k resolution accurate metric depth estimation.
\newblock \emph{arXiv preprint arXiv:2412.14015}, 2024.

\bibitem[Liu et~al.(2024{\natexlab{a}})Liu, Li, Chen, Li, Xu, and Plummer]{panofree}
Aoming Liu, Zhong Li, Zhang Chen, Nannan Li, Yi Xu, and Bryan~A Plummer.
\newblock Panofree: Tuning-free holistic multi-view image generation with cross-view self-guidance.
\newblock In \emph{European Conference on Computer Vision}, pages 146--164. Springer, 2024{\natexlab{a}}.

\bibitem[Liu et~al.(2023)Liu, Ren, Siarohin, Skorokhodov, Li, Lin, Liu, Liu, and Tulyakov]{hyperhuman}
Xian Liu, Jian Ren, Aliaksandr Siarohin, Ivan Skorokhodov, Yanyu Li, Dahua Lin, Xihui Liu, Ziwei Liu, and Sergey Tulyakov.
\newblock Hyperhuman: Hyper-realistic human generation with latent structural diffusion.
\newblock \emph{arXiv preprint arXiv:2310.08579}, 2023.

\bibitem[Liu et~al.(2024{\natexlab{b}})Liu, Cheng, Wang, Wang, Ouyang, Tan, Zhu, Shen, Chen, and Luo]{depthlab}
Zhiheng Liu, Ka~Leong Cheng, Qiuyu Wang, Shuzhe Wang, Hao Ouyang, Bin Tan, Kai Zhu, Yujun Shen, Qifeng Chen, and Ping Luo.
\newblock Depthlab: From partial to complete.
\newblock \emph{arXiv preprint arXiv:2412.18153}, 2024{\natexlab{b}}.

\bibitem[May and Aliaga(2023)]{cubegan}
Christopher May and Daniel Aliaga.
\newblock {CubeGAN: Omnidirectional Image Synthesis Using Generative Adversarial Networks}.
\newblock In \emph{Computer Graphics Forum}, pages 213--224. Wiley Online Library, 2023.

\bibitem[Peebles and Xie(2023)]{dit}
William Peebles and Saining Xie.
\newblock {Scalable Diffusion Models with Transformers}.
\newblock In \emph{Proceedings of the IEEE/CVF international conference on computer vision}, pages 4195--4205, 2023.

\bibitem[Persson(accessed 02/2025)]{humus}
Emil Persson.
\newblock Texture from {Humus}.
\newblock \emph{https://www.humus.name/index.php?page=Textures}, accessed 02/2025.

\bibitem[Podell et~al.(2024)Podell, English, Lacey, Blattmann, Dockhorn, M{\"u}ller, Penna, and Rombach]{sdxl}
Dustin Podell, Zion English, Kyle Lacey, Andreas Blattmann, Tim Dockhorn, Jonas M{\"u}ller, Joe Penna, and Robin Rombach.
\newblock {SDXL: Improving Latent Diffusion Models for High-Resolution Image Synthesis}.
\newblock In \emph{International Conference on Learning Representations}, 2024.

\bibitem[polyhaven.com(accessed 02/2025)]{polyhaven}
polyhaven.com.
\newblock {HDRIs}.
\newblock \emph{https://polyhaven.com/hdris}, accessed 02/2025.

\bibitem[Rombach et~al.(2022)Rombach, Blattmann, Lorenz, Esser, and Ommer]{latentdiffusion}
Robin Rombach, Andreas Blattmann, Dominik Lorenz, Patrick Esser, and Bj{\"o}rn Ommer.
\newblock {High-Resolution Image Synthesis with Latent Diffusion Models}.
\newblock In \emph{Proceedings of the IEEE/CVF Conference on Computer Vision and Pattern Recognition}, pages 10684--10695, 2022.

\bibitem[Salimans and Ho(2022)]{vprediction}
Tim Salimans and Jonathan Ho.
\newblock {Progressive Distillation for Fast Sampling of Diffusion Models}.
\newblock In \emph{International Conference on Learning Representations}, 2022.

\bibitem[Salimans et~al.(2016)Salimans, Goodfellow, Zaremba, Cheung, Radford, and Chen]{salimans2016improved}
Tim Salimans, Ian Goodfellow, Wojciech Zaremba, Vicki Cheung, Alec Radford, and Xi Chen.
\newblock Improved techniques for training gans.
\newblock \emph{Advances in neural information processing systems}, 29, 2016.

\bibitem[Saxena et~al.(2023{\natexlab{a}})Saxena, Herrmann, Hur, Kar, Norouzi, Sun, and Fleet]{ddvm}
Saurabh Saxena, Charles Herrmann, Junhwa Hur, Abhishek Kar, Mohammad Norouzi, Deqing Sun, and David~J. Fleet.
\newblock The surprising effectiveness of diffusion models for optical flow and monocular depth estimation.
\newblock In \emph{Thirty-seventh Conference on Neural Information Processing Systems}, 2023{\natexlab{a}}.

\bibitem[Saxena et~al.(2023{\natexlab{b}})Saxena, Kar, Norouzi, and Fleet]{depthgen}
Saurabh Saxena, Abhishek Kar, Mohammad Norouzi, and David~J Fleet.
\newblock Monocular depth estimation using diffusion models.
\newblock \emph{arXiv preprint arXiv:2302.14816}, 2023{\natexlab{b}}.

\bibitem[Shi et~al.(2024)Shi, Wang, Ye, Mai, Li, and Yang]{mvdream}
Yichun Shi, Peng Wang, Jianglong Ye, Long Mai, Kejie Li, and Xiao Yang.
\newblock {MVDream: Multi-view Diffusion for 3D Generation}.
\newblock In \emph{International Conference on Learning Representations}, 2024.

\bibitem[Somanath and Kurz(2021)]{somanath2021hdr}
Gowri Somanath and Daniel Kurz.
\newblock Hdr environment map estimation for real-time augmented reality.
\newblock In \emph{Proceedings of the IEEE/CVF conference on computer vision and pattern recognition}, pages 11298--11306, 2021.

\bibitem[Song et~al.(2021)Song, Meng, and Ermon]{ddim}
Jiaming Song, Chenlin Meng, and Stefano Ermon.
\newblock {Denoising Diffusion Implicit Models}.
\newblock In \emph{International Conference on Learning Representations}, 2021.

\bibitem[Stan et~al.(2023{\natexlab{a}})Stan, Wofk, Aflalo, Tseng, Cai, Paulitsch, and Lal]{ldm3dvr}
Gabriela Ben~Melech Stan, Diana Wofk, Estelle Aflalo, Shao-Yen Tseng, Zhipeng Cai, Michael Paulitsch, and Vasudev Lal.
\newblock {LDM3D-VR: Latent Diffusion Model for 3D VR}.
\newblock \emph{arXiv preprint arXiv:2311.03226}, 2023{\natexlab{a}}.

\bibitem[Stan et~al.(2023{\natexlab{b}})Stan, Wofk, Fox, Redden, Saxton, Yu, Aflalo, Tseng, Nonato, Muller, et~al.]{ldm3d}
Gabriela Ben~Melech Stan, Diana Wofk, Scottie Fox, Alex Redden, Will Saxton, Jean Yu, Estelle Aflalo, Shao-Yen Tseng, Fabio Nonato, Matthias Muller, et~al.
\newblock {LDM3D: Latent Diffusion Model for 3D}.
\newblock \emph{arXiv preprint arXiv:2305.10853}, 2023{\natexlab{b}}.

\bibitem[Tang et~al.(2023)Tang, Zhang, Chen, Wang, and Furukawa]{mvdiffusion}
Shitao Tang, Fuyang Zhang, Jiacheng Chen, Peng Wang, and Yasutaka Furukawa.
\newblock {MVDiffusion: Enabling Holistic Multi-view Image Generation with Correspondence-Aware Diffusion}.
\newblock In \emph{Proceedings of the International Conference on Neural Information Processing Systems}, 2023.

\bibitem[Wang et~al.(2024)Wang, Xiang, Fan, and Xue]{stitchdiffusion}
Hai Wang, Xiaoyu Xiang, Yuchen Fan, and Jing-Hao Xue.
\newblock Customizing 360-degree panoramas through text-to-image diffusion models.
\newblock In \emph{Proceedings of the IEEE/CVF Winter Conference on Applications of Computer Vision}, pages 4933--4943, 2024.

\bibitem[Wang et~al.(2023)Wang, Chen, Ling, Xie, and Song]{panodiff}
Jionghao Wang, Ziyu Chen, Jun Ling, Rong Xie, and Li Song.
\newblock 360-degree panorama generation from few unregistered nfov images.
\newblock In \emph{Proceedings of the 31st ACM International Conference on Multimedia}, pages 6811--6821, 2023.

\bibitem[Wang and Liu(2024)]{depthanywhere}
Ning-Hsu~Albert Wang and Yu-Lun Liu.
\newblock {Depth Anywhere: Enhancing 360 Monocular Depth Estimation via Perspective Distillation and Unlabeled Data Augmentation}.
\newblock \emph{Advances in Neural Information Processing Systems}, 37:\penalty0 127739--127764, 2024.

\bibitem[Wu et~al.(2024)Wu, Zheng, and Cham]{panodiffusion}
Tianhao Wu, Chuanxia Zheng, and Tat-Jen Cham.
\newblock {PanoDiffusion: 360-degree Panorama Outpainting via Diffusion}.
\newblock In \emph{International Conference on Learning Representations}, 2024.

\bibitem[Xiao et~al.(2012)Xiao, Ehinger, Oliva, and Torralba]{xiao2012recognizing}
Jianxiong Xiao, Krista~A Ehinger, Aude Oliva, and Antonio Torralba.
\newblock Recognizing scene viewpoint using panoramic place representation.
\newblock In \emph{2012 IEEE conference on computer vision and pattern recognition}, pages 2695--2702. IEEE, 2012.

\bibitem[Xu et~al.(2018)Xu, Ouyang, Wang, and Sebe]{xu2018pad}
Dan Xu, Wanli Ouyang, Xiaogang Wang, and Nicu Sebe.
\newblock Pad-net: Multi-tasks guided prediction-and-distillation network for simultaneous depth estimation and scene parsing.
\newblock In \emph{Proceedings of the IEEE conference on computer vision and pattern recognition}, pages 675--684, 2018.

\bibitem[Yang et~al.(2024{\natexlab{a}})Yang, Dong, Ma, Hu, Liu, Cui, and Ma]{yang2024dreamspace}
Bangbang Yang, Wenqi Dong, Lin Ma, Wenbo Hu, Xiao Liu, Zhaopeng Cui, and Yuewen Ma.
\newblock Dreamspace: Dreaming your room space with text-driven panoramic texture propagation.
\newblock In \emph{2024 IEEE Conference Virtual Reality and 3D User Interfaces (VR)}, pages 650--660. IEEE, 2024{\natexlab{a}}.

\bibitem[Yang et~al.(2024{\natexlab{b}})Yang, Kang, Huang, Zhao, Xu, Feng, and Zhao]{depthanythingv2}
Lihe Yang, Bingyi Kang, Zilong Huang, Zhen Zhao, Xiaogang Xu, Jiashi Feng, and Hengshuang Zhao.
\newblock {Depth Anything V2}.
\newblock \emph{Advances in Neural Information Processing Systems}, 37:\penalty0 21875--21911, 2024{\natexlab{b}}.

\bibitem[Ye et~al.(2024)Ye, Ji, Chen, Gao, Huang, Zhang, Ouyang, He, Zhao, and Zhang]{diffpano}
Weicai Ye, Chenhao Ji, Zheng Chen, Junyao Gao, Xiaoshui Huang, Song-Hai Zhang, Wanli Ouyang, Tong He, Cairong Zhao, and Guofeng Zhang.
\newblock Diffpano: Scalable and consistent text to panorama generation with spherical epipolar-aware diffusion.
\newblock \emph{arXiv preprint arXiv:2410.24203}, 2024.

\bibitem[Yu et~al.(2024)Yu, Duan, Herrmann, Freeman, and Wu]{wonderworld}
Hong-Xing Yu, Haoyi Duan, Charles Herrmann, William~T Freeman, and Jiajun Wu.
\newblock Wonderworld: Interactive 3d scene generation from a single image.
\newblock \emph{arXiv preprint arXiv:2406.09394}, 2024.

\bibitem[Zhang et~al.(2024)Zhang, Wu, Gambardella, Huang, Phung, Ouyang, and Cai]{panfusion}
Cheng Zhang, Qianyi Wu, Camilo~Cruz Gambardella, Xiaoshui Huang, Dinh Phung, Wanli Ouyang, and Jianfei Cai.
\newblock {Taming Stable Diffusion for Text to 360 Panorama Image Generation}.
\newblock In \emph{Proceedings of the IEEE/CVF Conference on Computer Vision and Pattern Recognition}, pages 6347--6357, 2024.

\bibitem[Zhao et~al.(2023)Zhao, Rao, Liu, Liu, Zhou, and Lu]{vpd}
Wenliang Zhao, Yongming Rao, Zuyan Liu, Benlin Liu, Jie Zhou, and Jiwen Lu.
\newblock Unleashing text-to-image diffusion models for visual perception.
\newblock In \emph{Proceedings of the IEEE/CVF International Conference on Computer Vision}, pages 5729--5739, 2023.

\bibitem[Zheng et~al.(2020)Zheng, Zhang, Li, Tang, Gao, and Zhou]{structured3d}
Jia Zheng, Junfei Zhang, Jing Li, Rui Tang, Shenghua Gao, and Zihan Zhou.
\newblock {Structured3D: A Large Photo-Realistic Dataset for Structured 3D Modeling}.
\newblock In \emph{Computer Vision--ECCV 2020: 16th European Conference, Glasgow, UK, August 23--28, 2020, Proceedings, Part IX 16}, pages 519--535. Springer, 2020.

\bibitem[Zhu et~al.(2024)Zhu, Wang, Huang, Ye, Ouyang, and He]{zhu2024point}
Haoyi Zhu, Yating Wang, Di Huang, Weicai Ye, Wanli Ouyang, and Tong He.
\newblock Point cloud matters: Rethinking the impact of different observation spaces on robot learning.
\newblock \emph{Advances in Neural Information Processing Systems}, 37:\penalty0 77799--77830, 2024.

\end{thebibliography}
